\documentclass[twocolumn,aps,floats,nofootinbib,prd,amsmath,amssymb,secnumarabic]{revtex4}
\usepackage{graphicx, epsfig, natbib, color}

\newcommand{\etal}{et al.}
\newcommand{\BST}{Boyle \etal\ \cite{Boyle:2005ug}}

\newcommand{\phisxty}{\phi_{60}}
\newcommand{\phim}{\phi_{m}}

\begin{document}

\title{Fine-tuning criteria for inflation and the search for primordial gravitational waves}
\author{Simeon Bird$^1$\footnote{spb41@cam.ac.uk}, Hiranya V. Peiris$^1$, Richard Easther$^2$} 
\affiliation{$^1$Institute of Astronomy, University of Cambridge, Cambridge CB3 0HA, U.K.\\
$^2$Department of Physics, Yale University, New Haven, CT 06520, U.S.A.}

\begin{abstract}
We revisit arguments that simple models of inflation with a small red tilt in the scalar power spectrum generically yield an observable tensor spectrum.
We show that criteria for fine-tuning based upon the algebraic simplicity of the potential depend strongly upon the explicit assumptions they incorporate, particularly regarding the end of inflation. 
In addition, some models with algebraically simple potentials require carefully tuned initial field configurations, and not all types of fine-tuning are identifiable via the algebraic simplicity of the potential. Conversely, in the absence of a strong prior on the mechanism that ends inflation, we demonstrate the existence of potentials with vanishingly small tensor amplitudes which are natural in terms of both their algebraic form and initial conditions. We thus argue that proposed experiments (CMBPol or BBO) which make highly sensitive measurements of the tensor amplitude cannot definitively rule out the inflationary paradigm. 
\end{abstract}

\maketitle

\section{Introduction}

 In the simplest scenarios, inflation is driven by a single scalar field, the ``inflaton'', rolling in a slowly varying potential  \cite{Guth:1981,Sato:1981,Mukhanov:1981,Linde:1982,Albrecht:1982}. Sufficient inflation ($\approx 60$ $e$-folds) smooths the universe and solves the flatness, horizon, and monopole problems. Initial inhomogeneities originate as quantum fluctuations in the inflaton field and the metric, sourcing scalar \cite{Hawking:1982,Guth:1982, Starobinsky:1982, Bardeen:1983} and tensor \cite{Grishchuk:1975, Starobinsky:1979} perturbations which are almost Gaussian and scale-invariant \cite{Bardeen:1983}. These broad-brush predictions are consistent with observations  (e.g. \cite{Komatsu:2008hk,Peiris:2006ug,Peiris:2006sj,Peiris:2008be}). 

The shape of the inflaton potential is not predicted by theory; it is only required to satisfy the slow-roll conditions. However, the shape of the potential is constrained by the scalar spectral index, $n_s$, which measures the deviation of the scalar power spectrum from scale invariance, and the tensor-scalar ratio, $r$, which measures the amplitude of the tensor modes relative to that of the scalar modes. Data from the {\sl Wilkinson Microwave Anisotropy Probe} (WMAP) combined with a compilation of supernovae and baryon acoustic oscillations data gives $n_s =0.960^{+0.014}_{-0.013}$ at 68\% CL if the tensor-scalar ratio $r=0$, and $n_s=0.968 \pm 0.015$ when we marginalize over $r$  \cite{Dunkley:2008ie, Komatsu:2008hk}. There is currently no evidence for the existence of a primordial tensor background: the 95\% CL upper limit from the same data compilation is $r<0.20$.

A detection of primordial tensors would have tremendous implications for cosmology as their existence is a key prediction of inflation \cite{Starobinsky:1979}. Moreover, measuring a non-zero value of $r$ would eliminate the ekpyrotic/cyclic scenarios, which predict an unobservably small primordial tensor amplitude \cite{Khoury:2001bz,Boyle:2003km}.
Finally, the observed amplitude of the tensor spectrum fixes the energy scale of inflation, removing a major uncertainty in inflationary model building. On the other hand, an observable tensor spectrum would imply the need for super-Planckian field variations during inflation in simple single field models \cite{Lyth:1996im}, which raises specific theoretical challenges.
 
Cosmic variance, polarized foregrounds, and weak lensing of the $E$-mode may ensure that $r \sim 10^{-3}$ is the effective lower limit for a detection of tensor modes in the cosmic microwave background (CMB) (e.g. \cite{Knox:2002pe,Kesden:2002,Verde:2005ff,Amarie:2005in}). Consequently, it is important to know whether inflation makes a {\em generic\/} prediction for the value of $r$ and, if so, whether it falls within the detectable range. A plethora of explicit inflationary models with varying degrees of physical motivation have been proposed. Many models do have observable values of $r$ (e.g. chaotic inflation \cite{Linde:1983}); conversely, a detection of tensors would rule out many \cite{Baumann:2006cd}, but not all \cite{Dimopoulos:2005ac,Easther:2005zr,Silverstein:2008}, explicit string-theory constructions. Consequently, to show that inflation has a ``natural'' range of $r$ values, we must establish a ``weight'' that will give preference to certain classes of models.

One weighting proposal is that of \BST\ (BST), who offer a fine-tuning statistic based on the algebraic simplicity of the potential,  
along with five conditions which a successful inflationary potential should satisfy. These criteria are then applied to quartic polynomial potentials, with the conclusion that a high degree of fine-tuning is required for $n_s>0.98$, or for $r<10^{-2}$ with $n_s>0.95$. If future observations confirm that $0.95 < n_s< 0.98$, the BST criterion suggests that $r$ is within the observable range.

In addition to tuning the {\em parameters\/} of the potential, some potentials need a very special initial field configuration if inflation is to begin. This tuning can involve both the suppression of the kinetic energy relative to the potential energy  (to ensure that the inflaton does not ``overshoot'' the inflationary portion of the potential) or the absence of inhomogeneities on the initial inflationary patch. This topic has been investigated for many of the common models of inflation \cite{Belinsky:1985, Goldwirth:1990, Mendes:2000,Underwood:2008dh, Chiba:2008}.

In particle physics, models are regarded as fine-tuned if they require correlations between parameters beyond those imposed by the symmetries of the underlying theory \cite{tHooft:1979bh}, or when certain parameters are arbitrarily set to zero; models which are not fine-tuned in this sense are known as ``technically natural'' \cite{Hotchkiss:2008sa}. 

There are other definitions of naturalness based on the sensitivity of observables to small changes in their parameters \cite{Barbieri:1988, Anderson:1994dz}. Ref. \cite{Hotchkiss:2008sa} examined technically natural potentials which did not obey all of the BST viability conditions, and found red-tilted examples with $r<0.01$ which were not fine-tuned under their definition. Other studies came to similar conclusions \cite{Efstathiou:2006, Smith:2008pf}. Inflationary potentials are often written down without reference to an underlying theory, so we do not necessarily know the relevant symmetries, nor the couplings between the inflaton and other fields which contribute loop corrections to the inflaton potential.   
Consequently, while technical naturalness is a powerful tool, it is less useful when applied to generic inflationary potentials. Moreover, inflationary potentials typically require some tuning in order that inflation happens at all.

We tackle two questions. Firstly, we show that even a mild relaxation of the rules imposed by BST significantly weakens their conclusions for the likely range of $r$. In addition, we ask whether the inflaton's initial conditions must be tuned in order for viable inflation to occur. We consider only the ratio of the kinetic to potential energy, treating the universe as a homogeneous system and ignoring the ability of primordial inhomogeneities to suppress the onset of inflation. However, this requirement places a strong constraint on the inflationary model space, which does not overlap with the BST proposal, demonstrating the interplay between the {\em definition\/} of tuning, and the resulting conclusions for the likely values of $r$ and $n_s$. Our aim is not to assess the intrinsic merit of different naturalness criteria. Rather, our purpose is to investigate whether conclusions about the natural value of $r$ are sensitive to assumptions encoded in the tuning criteria. This question is of considerable importance when assessing and justifying CMB $B$-mode or gravitational wave experiments sensitive to $r$. Likewise, any conclusions we reach regarding the ``natural'' value of $r$ in inflation will determine the significance of a null result from these experiments.

\section{Methodology}\label{sec:methods}
 
\subsection{Notation}\label{sec:notation}

We use natural units, $c=\hbar=1$, and $M_P$ is the reduced Planck mass. The first two slow-roll parameters are
\begin{eqnarray}
\epsilon &=& \frac{M^2_P}{2}\left[\frac{V'(\phi)}{V(\phi)}\right]^2 \, ,\\
\eta &=& M^2_P\left[\frac{V''(\phi)}{V(\phi)} \right] \, ,
\end{eqnarray}
where $V(\phi)$ is the inflaton potential, $\phi$ is the inflaton and primes denote derivatives with respect to $\phi$. To first order, the spectral index and the scalar-tensor ratio are
\begin{eqnarray}
n_s &=& 1 + 2\eta-6\epsilon  \, ,\\
r &=& 16\epsilon \, .
\end{eqnarray}
These quantities are evaluated $60$ $e$-folds before the end of inflation, $\phi_e$, when $\phi=\phi_{60}$, which is taken to correspond to the CMB scales\footnote{Our conclusions are not sensitive to the details of reheating.}. A minimally coupled inflaton with a canonical kinetic term has the equation of motion
\begin{equation}
\ddot{\phi}+3H\dot{\phi}+V'(\phi) = 0\,,
\end{equation}
where $H$ is the Hubble parameter. The scale factor $a(t)$ is described by
\begin{equation}
H^2 = \left(\frac{\dot{a}}{a}\right)^2 = \frac{1}{3M_P^2}\left[V(\phi)+\frac{1}{2}\dot{\phi}^2\right]\, ,
\end{equation}
where the overdot denotes a derivative with respect to coordinate time $t$.
It is convenient to use the number of $e$-folds, $N$, as the time variable, via
\begin{equation}
dN = H dt \, .
\end{equation}
Hence both $N$ and $\phi$ increase as $t$ increases\footnote{The opposite sign convention can be found in the literature; with consistent usage, there is no practical distinction.}. Defining $\dot{\phi} = \Pi$, the equations of motion are
\begin{eqnarray}
\frac{d\phi}{dN} &=& \frac{\Pi}{H} \,,\\
\frac{d\Pi}{dN} &=& -\left(3\Pi +\frac{V'}{H}\right) \,.
\end{eqnarray}
Provided $\eta-\epsilon \ll 1$, an attractor solution exists for this system \cite{Goldwirth:1990}, satisfying
\begin{eqnarray}
\frac{d\Pi}{dN} = 0 \;  \implies \; \Pi = -\frac{V'}{3H}\,.
\end{eqnarray}
In every result quoted in this paper, we have solved the full equations of motion numerically; no slow-roll approximation has been employed.

\subsection{Selection Conditions} \label{selection}

 Following BST \cite{Boyle:2005ug}, we consider quartic polynomial potentials and apply five conditions which define a viable inflationary model. These conditions largely overlap with those of BST; since we are investigating the impact of {\em changing\/} the viability conditions we specify them in some detail:
\begin{enumerate}
\item The density perturbations have an amplitude of roughly $10^{-5}$ $60$ $e$-folds before the end of inflation, in agreement with observations.
\label{cond:perts}
 \item At least $60$ $e$-folds of inflation occur, with the trajectory starting on the slow-roll inflationary attractor and $\phi$ evolves monotonically as a function of time. 
\label{cond:e-folds}
\item Inflation terminates in a deterministic and smooth manner, via $\epsilon=1$, rather than through a hybrid-type transition \cite{Linde:1993cn}.\label{cond:end}
\item After inflation, the field evolves smoothly to a minimum of the potential located at $\phi_m$, and $V(\phi_m)=0$.\label{cond:min}
\item $V$ is bounded below, and the minimum, if metastable, is long-lived. 
\label{cond:meta}
\end{enumerate}

Beyond specifying these rules, we must also write down a generic potential. We follow BST and write down the quartic polynomial\footnote{The inflaton potential is the {\em effective\/} potential of $\phi$ and thus includes all loop corrections. Consequently, it can include terms that would be non-renormalizable if they appeared at tree level, and the restriction to the quartic form is effectively a further prior based on algebraic simplicity. Note that this cut excludes potentials of the Coleman-Weinberg form \cite{Albrecht:1982,Linde:1983a}.}
 
\begin{equation} \label{generic}
V(\phi) = V_0(\phi^4+B\phi^3+A\phi^2+C\phi+D)\,.
\end{equation}

To satisfy condition (\ref{cond:min}), we require a minimum at $\phi_m$, with $V(\phi_m)=0$. Without loss of generality, we can set $\phi_m=0$, which implies that $C=D=0$. To satisfy condition (\ref{cond:meta}), the potential must be bounded below, which implies $V_0 \geq 0$. The magnitude of $V_0$ is dynamically irrelevant, so we can choose it such that condition (\ref{cond:perts}) is satisfied. Typical values of $V_0$ were around $10^{-16}$; in no case did one of our tested potentials exceed $V(\phi) = M^4_P$. This is necessary \cite{Linde:1983} (but not sufficient) to keep quantum gravity effects under control. Our potential now has two free parameters, 
\begin{equation}
V = V_0\phi^2(\phi^2+B\phi+A) \, .\label{eq:eftpotential}
\end{equation}
We set $A \geq 0$ to ensure that the extremum at the origin is a minimum\footnote{For computational reasons, we also impose $A<10^5$, but this makes no difference to our results.}. We apply a slightly stricter version of condition (\ref{cond:meta}), namely that the minimum is {\em stable\/}, eliminating ambiguity in the definition of ``long-lived''. Now we can impose $V(\phi) \geq 0$, since the inflaton must evolve toward the lowest-lying point of the potential, which is assumed to be at $V(0)=0$ (via condition (\ref{cond:end})). Hence $V(\phi)=0$ has at most two solutions for a valid potential, one of which is at the origin, so that $\phi^2 + B\phi + A = 0$ has at most one solution, for which we need an imaginary or vanishing determinant and thus
\begin{equation}
-2\sqrt{A} \leq B \leq 2\sqrt{A} \, .
\end{equation}

We construct candidate potentials by randomly sampling $A$ and $B$ with logarithmic priors within the given bounds and numerically testing to determine whether conditions (\ref{cond:e-folds}) and (\ref{cond:end}) are met. After imposing these rules, we can only populate a subset of the $(n_s,r)$ plane, as the assumed functional form of the potential and the selection conditions constitute a strong theoretical prior on the slow-roll parameters, and hence the observables. 
Such restrictive theoretical priors can lead to a "detection" of a tensor-scalar ratio even when the likelihood with respect to the data shows no evidence for a deviation from $r=0$ \cite{Destri:2008}. This point has also been made in \cite{Valkenburg:2008}.
Our results are in agreement with refs. \cite{Hoffman:2000ue, Peiris:2003ff, Boyle:2005ug}. We stress that this cut is independent of any ``weighting'' function proposed to assess the degree of tuning of potentials that do comply with the above conditions.

\subsection{Fine-Tuning Criteria}\label{sec:criteria}

We will discuss two fine-tuning criteria. The first is that proposed by BST, which measures the fine-tuning in the parameters of the potential. Our second measures fine-tuning in the initial configuration of field space, as shown by the sensitivity to an initial kinetic term.

\subsubsection{BST Fine-tuning Criterion}\label{sec:BST}

The fine-tuning criterion proposed by BST counts the number of ``unnecessary features'' occurring in the potential during the last $60$ $e$-folds of inflation ($Z_\eta$). A feature is a zero of some derivative of $\eta$ with respect to $\phi$, so
\begin{equation}
Z_\eta =  \sum_{p=0}^{P} \chi\left(\frac{d^p\eta}{d\phi^p}\right)\,, 
\end{equation}
where $\chi(f)$ counts the number of zeros of $f$ within a given range; in this case the last $60$ $e$-folds of inflation. Since all derivatives of $\eta$ may be non-trivial, the value of $P$ is arbitrary; we take $P=10$. Our implementation differs slightly from that of BST, who take $\chi(\eta_{,N})$ in the place of $\chi(\eta_{,\phi})$: derivatives of $N$ can introduce artificial features into the potential (via the $\phi_{,N}$ term), over-estimating the amount of fine-tuning. Our change weakens BST's conclusion that $n_s <0.98$ to $n_s <0.99$ for non-fine-tuned potentials, but has no other impact. 
BST stipulate that models with $Z_\eta > 1$ are fine-tuned, but their criterion is not intended to quantify the degree of fine-tuning in such cases; models with (say) $Z_\eta =12$ and $Z_\eta = 30$ would be placed on the same footing. Our graphs therefore show all points with $Z_\eta \geq 12 $ as equivalent.  

\subsubsection{Initial Conditions Fine-tuning Criterion}\label{sec:phasespace}

The question of whether inflation itself requires a fine-tuned initial field configuration has been discussed many times  (e.g. \cite{Belinsky:1985, Piran:1985, Belinsky:1986, Piran:1986, Gibbons:1987, Albrecht:1987, Belinsky:1988, Kung:1989, Kung:1990, Goldwirth:1990, Goldwirth:1990a, Goldwirth:1991, Goldwirth:1992}). Our alternative fine-tuning criterion examines the effect of a substantial initial kinetic term on classical inflationary trajectories in a flat, homogeneous universe.
This is similar to a criterion applied to chaotic inflation in refs. \cite{Belinsky:1985, Piran:1985} and to new inflation in ref. \cite{Goldwirth:1990}. More generally, one can ask whether inflation generically begins with initial conditions that are anisotropic, inhomogeneous or have non-zero spatial curvature \cite{Moss:1986, Albrecht:1987, Kung:1989, Kung:1990, Goldwirth:1990a, Goldwirth:1991}. We ignore these possibilities, but including them would in general strengthen the conclusions we obtain by working only with an initially arbitrary kinetic term.
The inflationary universe driven by a single scalar field is described by a second-order system of non-linear differential equations. We consider how sensitive this system is to its initial conditions, $\Pi_i$ and $\phi_i$, defining the criterion $R_i$ for a given potential as the fraction of initial conditions leading to valid inflationary trajectories. That is: 
\begin{equation}
R_i = \frac{\int{F(\phi_i,\Pi_i)d\phi_i d\Pi_i}}{\int d\phi_i d\Pi_i},
\end{equation}
where $F(\phi_i,\Pi_i)$ is an indicator function which is unity where ($\Pi_i$, $\phi_i$) lead to successful inflation, and zero otherwise. A model which is minimally fine-tuned has $R_i=1$, and one which is maximally fine-tuned has $R_i=0$. 

Notice that $d\phi_i d\Pi_i$ is not a dynamically invariant measure; the Hamiltonian phase space measure contains an extra factor $a^3(t)$. With this weighting, it has been shown that inflation is disfavored by $e^{-3N}$, where $N$ is the number of $e$-folds \cite{Gibbons:1987, Gibbons:2006pa} (for further discussion on whether inflation is generic, see also refs. \cite{Hollands:2002, Kofman:2002JHEP}). We avoid the complexities of the measure problem by assuming {\sl a priori} that inflation occurs, considering only the relative weighting of one inflationary model to another; we are not attempting to see whether inflation itself is natural, but instead to compare the degrees of naturalness of different inflationary potentials. 

To evaluate $F$, we must check whether the system satisfies conditions (\ref{cond:perts})--(\ref{cond:meta}) for a given $(\Pi_i,\;\phi_i)$. In some cases, inflation will end in a false vacuum -- a minimum other than $\phi=0$. These trajectories violate condition (\ref{cond:meta}) and so are discarded. In a flat, expanding, Friedmann-Robertson-Walker universe, the Hubble parameter is monotonic and fixes the total energy. We can use this to determine when the inflaton will be unable to cross a maximum in the potential without actually evolving the equations of motion. 
If the system expands through $300$ $e$-folds, yet the eventual destination is still indeterminate, because the total energy exceeds that at the maximum, we discard the trajectory. This situation is rare, and we impose this cut to avoid spending a disproportionate amount of effort evaluating a small set of trajectories. Finally, trajectories where $\phi$ is not monotonically increasing are discarded.

Computational limitations prevent us from considering every possible initial condition, so we limit ourselves to a small region $\sim 60$ $e$-folds before the end of inflation. The choice of the bounds on this region can change the numerical value of $R_i$ significantly. However, our aim is to compare different models, not to provide an absolute indicator of how fine-tuned they are, so we need only enough variation in $R_i$ to distinguish between them. 
To see how this works,  consider the possible reasons a given set of initial conditions can fail to yield sufficient inflation. If there is a feature in the potential (an extremum, or a point where $\epsilon=1$), most trajectories beyond it will be invalid. If the lower limit on $\phi$ is too close to zero, and a feature lies beyond it, a model would be unfairly favored. However, for large $|\phi|$, potentials are dominated by the $\phi^4$ term and so features of this kind are rare. In practice then, extending the range of $\phi$ will usually add a non-inflating region to potentials with such a feature, and an inflating region to potentials without one, thus accentuating the difference in $R_i$.

The other major cause of a failing trajectory is overshoot. This occurs when a large $\Pi_i$ pushes the field through the inflationary region too quickly, leaving insufficient time on the slow-roll attractor to acquire sufficient $e$-folds of expansion before the inflaton starts oscillating, or driving the field beyond the oscillatory region altogether. This problem is present to some extent in all potentials. 
When $V \lesssim \Pi^2/2$, the evolution is dominated by the kinetic term; in this region the effect of the potential is negligible, and so evolution does not serve to differentiate between models. We therefore chose as an upper bound for $\Pi$ the smallest value always within the kinetic dominated regime. We found that $V \lesssim 10^{-10}M_P^4$, so $\Pi_i < 10^{-4}M^2_P$. We also considered various other limits, with no changes in conclusions.

Since we ask that $\phi$ be increasing we need  $\Pi_i \geq 0$. For $\phi_i$, we take $\phi_i < \phi_{60}$, the point on the potential $60$-$e$-folds before the end of inflation, as measured by the attractor solution. It is the point where $n_s$ and $r$ are calculated, and defines the smallest value of $\phi$ from which a valid trajectory can start. Our upper limit on $\phi_i$ is $\phi_{60}+20M_P$, which is $\approx 2\phisxty$ for large-field models such as $V=\lambda\phi^4$. Our bounds are therefore:
\begin{eqnarray}
	\phi_{60} \leq &\phi_i & \leq (\phi_{60}+20M_P) \,,\\
0 \leq &\Pi_i & \leq 10^{-4}M^2_P\,. 
\end{eqnarray}
Finally, because $F$ is evaluated on a finite grid, we ensure that our mesh is fine enough to reduce the resulting error in $R_i$ to less than 1\%. 

\section{Results}\label{sec:results}
 
\begin{figure*}[!htp]
\includegraphics[scale=0.175]{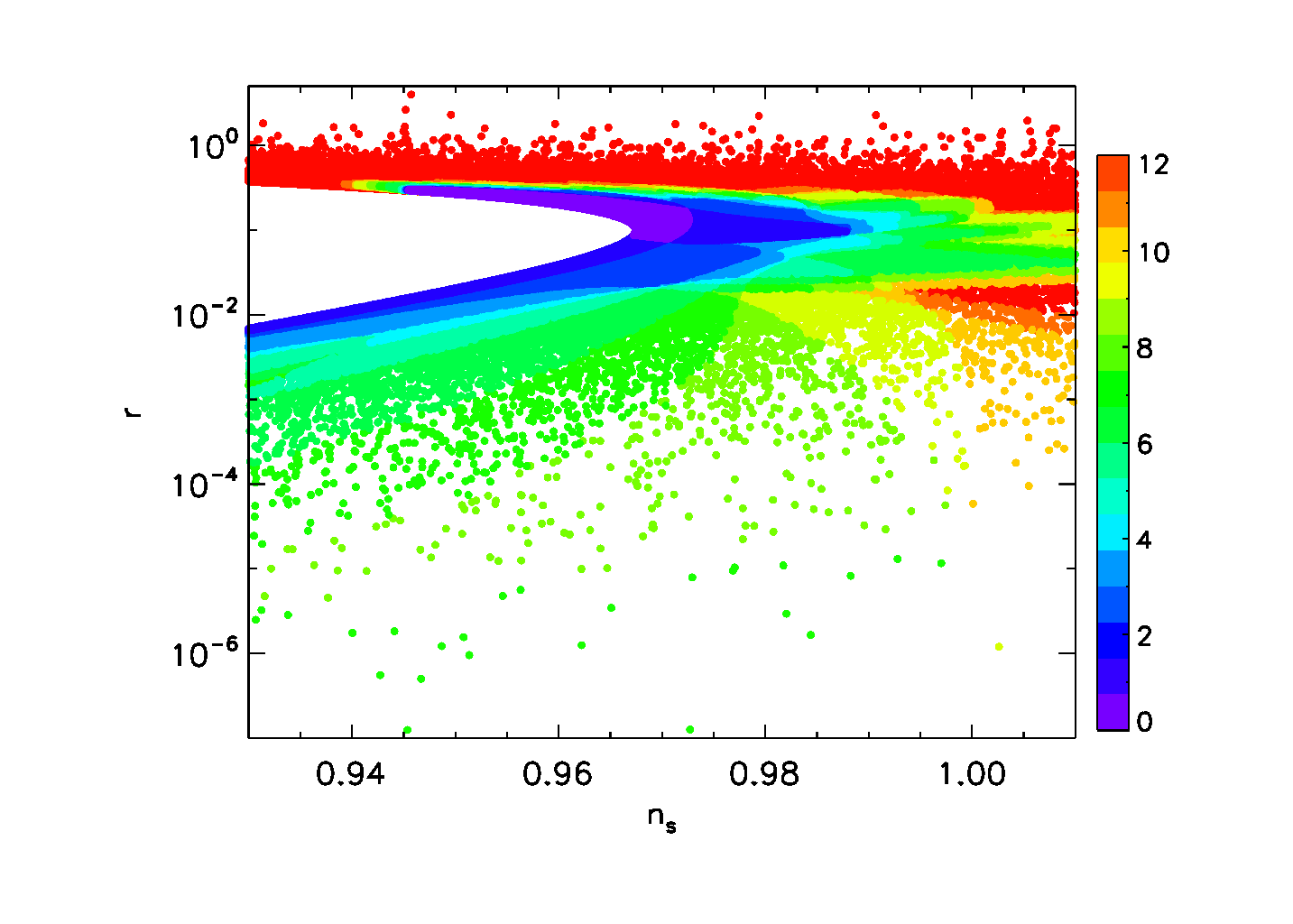} \hfill
\includegraphics[scale=0.175]{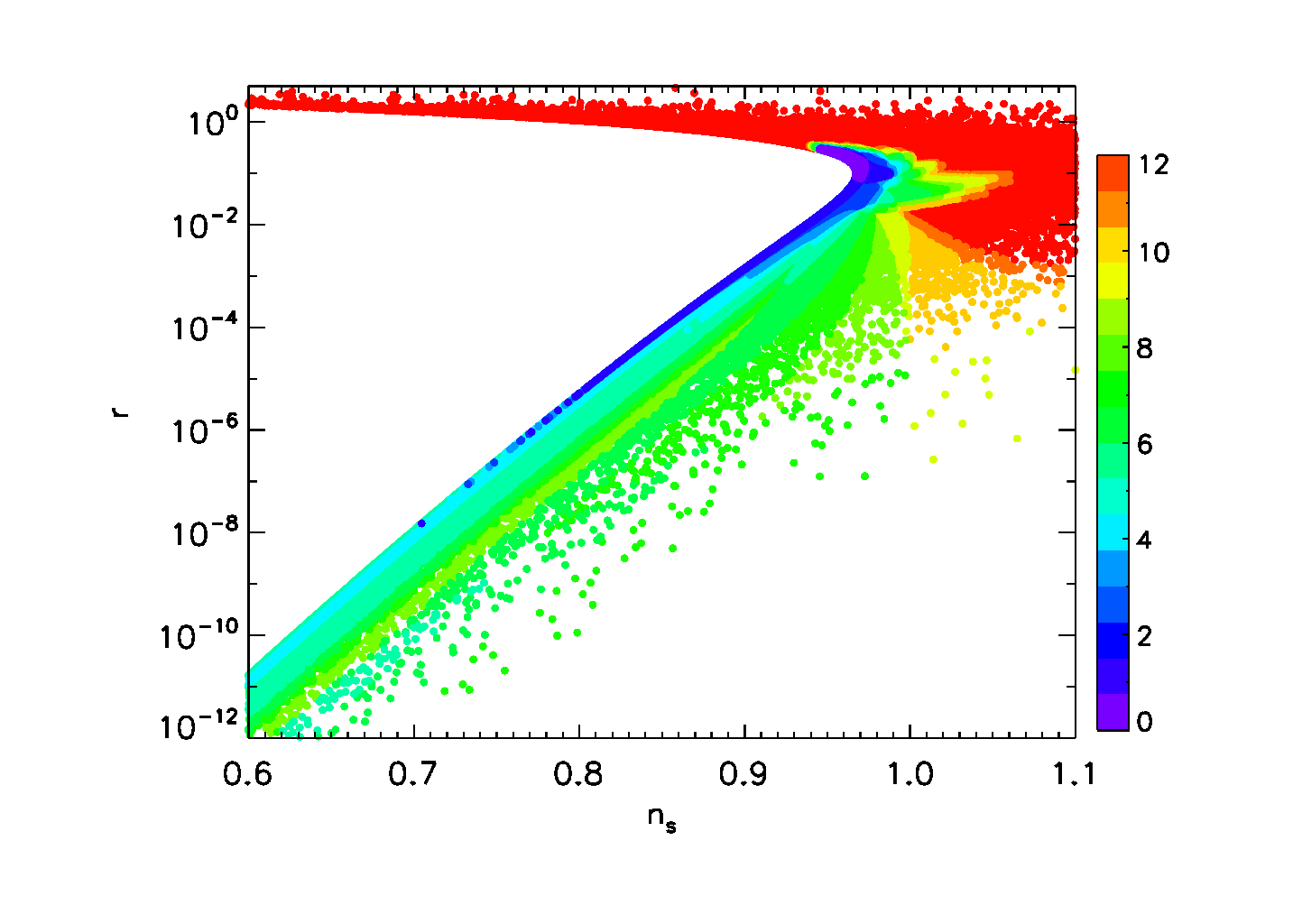}
\caption{The BST fine-tuning criterion with quartic potentials. The color scale shows the value of $Z_\eta$ in the $(n_s,r)$ plane. The models were selected using a logarithmic prior for $A$ and $B$. (Left) A limited range shown for comparison with BST. (Right) $Z_\eta$ in an extended range. The tail of points with low fine-tuning and low $r$ extends beyond this diagram. All points with $Z_\eta\geq 12$ are shown in the same color. See Appendix \ref{forbidden} for an explanation of the ``forbidden zone'' in the middle.} \label{boylegraph}
\end{figure*}

\begin{figure*}[!htp]
\includegraphics[scale=0.175]{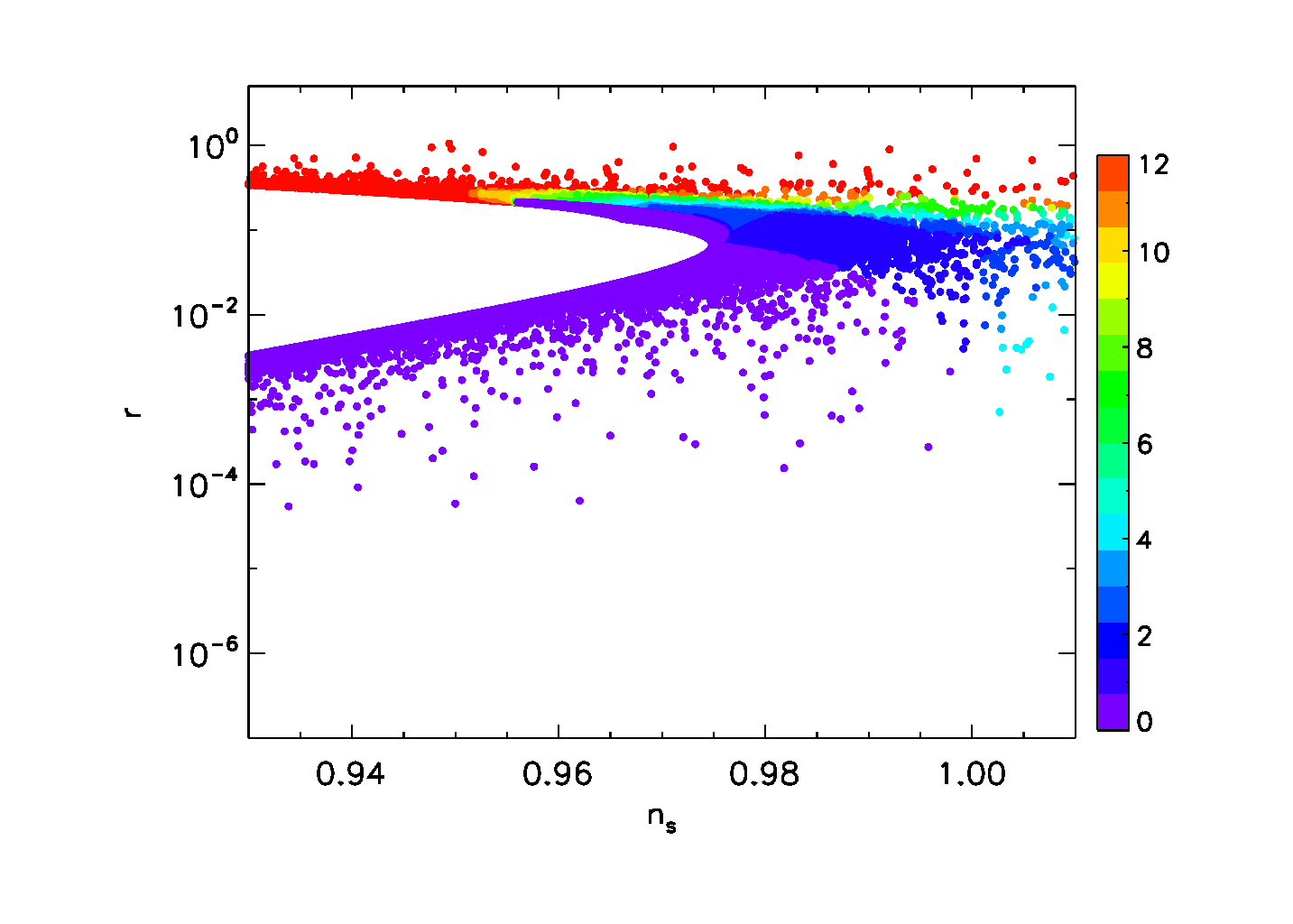} \hfill
\includegraphics[scale=0.175]{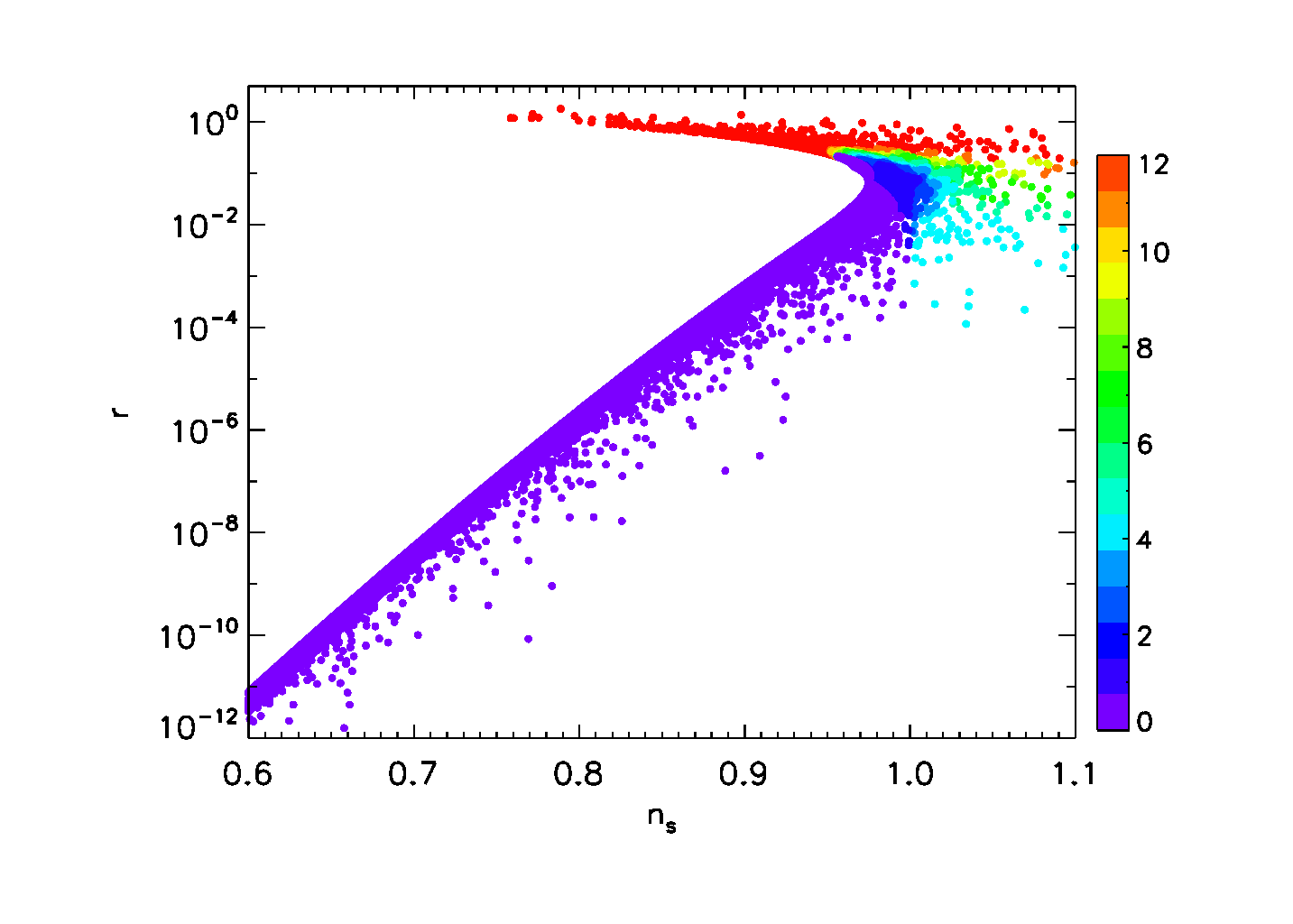}
\caption{The BST fine-tuning criterion with cubic potentials, relaxing selection conditions (\ref{cond:min}) and (\ref{cond:meta}). The color conventions are the same as those of Figure~\ref{boylegraph}. The right panel shows an extended range compared to the left.} \label{cubicgraph}
\end{figure*}

\begin{figure*}[!htp]
\includegraphics[scale=0.175]{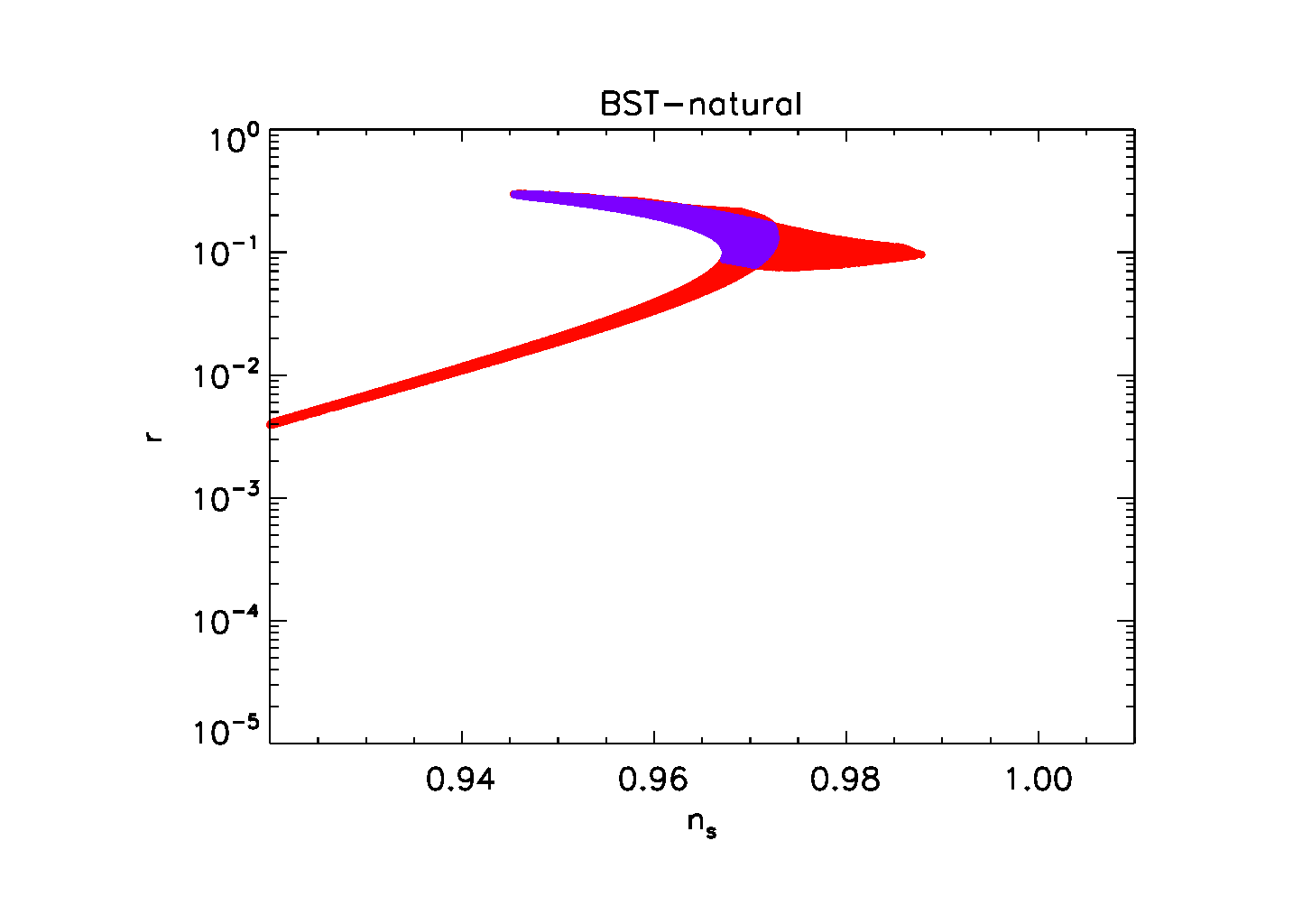} \hfill
\includegraphics[scale=0.175]{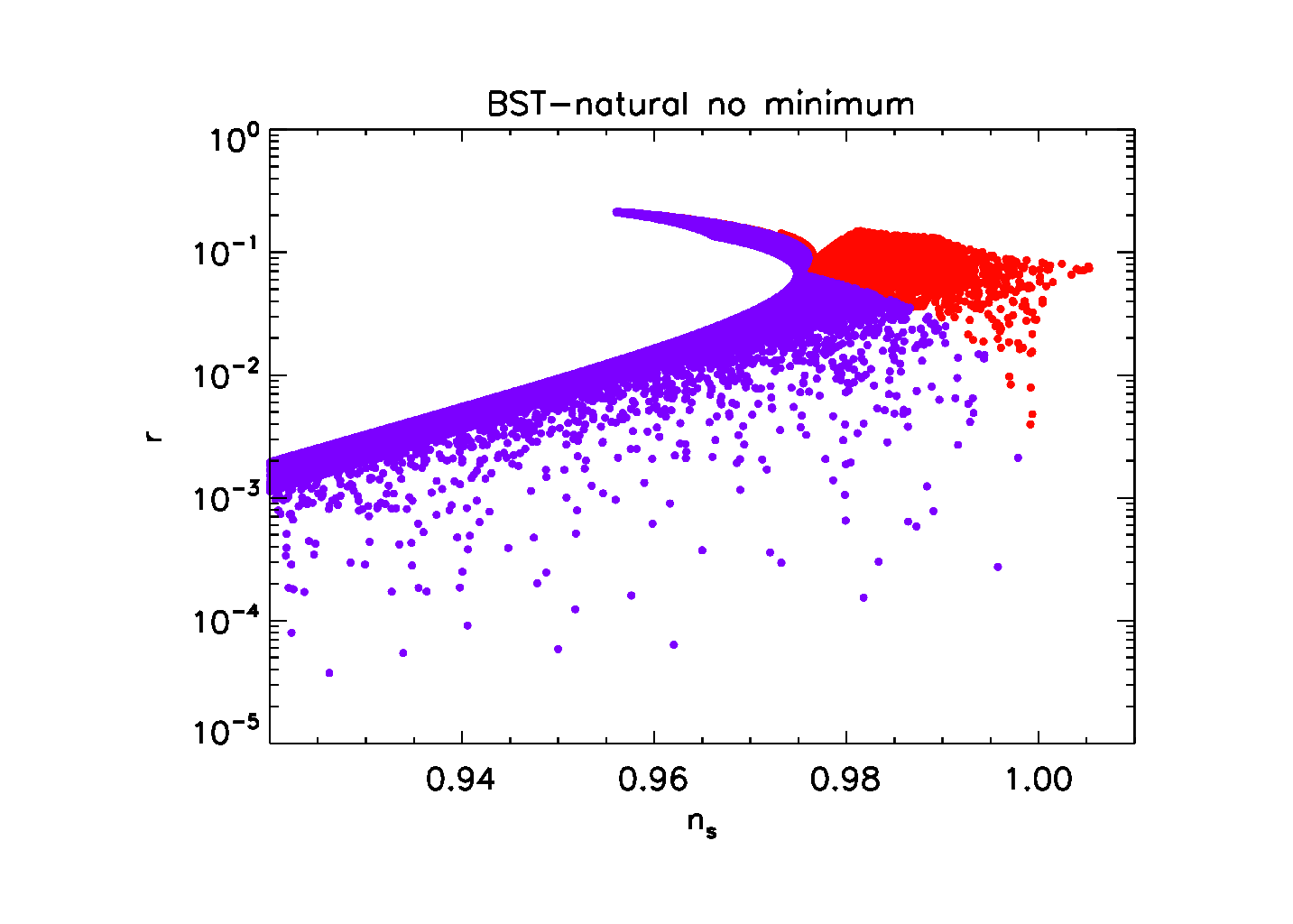}
\caption{Potentials not fine-tuned by the BST definition. Purple points have $Z_\eta=0$ and red points have $Z_\eta=1$. (Left) Quartic potentials satisfying all the BST selection conditions. (Right) Cubic potentials relaxing selection conditions (\ref{cond:min}) and (\ref{cond:meta}).} \label{boyle01graph}
\end{figure*}

\subsection{BST Fine-tuning Criterion}\label{sec:BSTresults}

As the left hand panel of Figure~\ref{boylegraph} shows, we recover Figure~1 of BST. The conclusion of BST that $n_s < 0.98$ has been weakened to $n_s < 0.99$; in addition, there are small differences in the shapes of some of the fine-tuning regions. These changes appear to be due to our use of $\phi$ derivatives instead of $N$ derivatives when defining $Z_\eta$, and will not affect the following discussion. 
The right hand panel of Figure~\ref{boylegraph} shows  a ``tail'' of untuned models with $r \ll 10^{-2}$, albeit at the cost of very red $n_s$, which is ruled out by data. This plot demonstrates that the BST criterion does not necessarily ``predict'' a {\sl small} deviation from scale-invariance, but rather that models with low $Z_\eta$ have correlated values of $n_s$ and $r$. If $n_s$ is at the lower limit of the currently permitted range ($\sim 0.94$), then $r\sim 0.01$, but if $n_s\sim 0.97$, $r>0.01$, and thus observable.
 
Figure~\ref{boylegraph} has regions with high $Z_\eta$ for $r>0.4$ and $r < 10^{-4}$. The corresponding potentials typically have a small region with decreased slope, or an inflection point. This is bracketed by two zeros of $V''$ (and thus $\eta$), which guarantees a high value of $Z_\eta$. The boost to $Z_\eta$ is less efficient in low-$r$ models because the inflaton moves more slowly and the second zero of $\eta$ is encountered more than 60 $e$-folds before the end of inflation.   
 
Having reproduced the essential features of BST's results, we now investigate what happens as the selection conditions are relaxed. First, the potential may have parameters which are irrelevant at the values of $\phi$ which determine $n_s$ and $r$, but which are important as inflation comes to an end. In this case, we do not expect conditions (\ref{cond:min}) and (\ref{cond:meta}) to hold, since we are only constraining the form of the potential as the cosmological perturbation spectrum is laid down. We also know that if we have a pure $\Lambda^4-(\phi/\mu)^4$ potential, $n_s\sim 0.95$ while $r$ can be very small. This potential has the form of equation (\ref{generic}) once we drop the positivity condition.  
We now relax conditions (\ref{cond:min}) and (\ref{cond:meta}), and consider cubic potentials with a linear term, which includes the ``inflection point'' potentials found in some string-theoretic constructions \cite{Baumann:2007np, Baumann:2007ah,Krause:2007jk, Linde:2007jn}, with the form
\begin{eqnarray}
V(\phi) = V_0(\phi^3 + A\phi^2+B\phi+C)\,.
\end{eqnarray}
Without loss of generality, we set $V(0)=0$, and $V_0 <0$, so that $V(\phi) \to -\infty$ as $\phi \to \infty$. Once condition (\ref{cond:min}) is abandoned, inflation is no longer tied to the deepest minimum, so there may be two regions satisfying the inflationary conditions, up to a rescaling of the potential so that the perturbation amplitude is correct. If there are two such regions, one will deposit the field in a (local) minimum, while the other will occur when $\phi$ is slowly rolling to infinity (or to $V\sim M_P^4$). In this case we include points for  $n_s$ and $r$ obtained from  both regions, and show our results in Figure~\ref{cubicgraph}.
Strikingly, there are now non-fine-tuned points with a slightly red $n_s$ and with $r<10^{-4}$. These points occur in a region of the cubic potential with no following minimum.

The whole upper arm of Figure~\ref{cubicgraph} contains points where, after inflation finishes, the inflaton ends in a minimum, while the lower arm contains points where it does not. The sharp delineation and curved feature at $n_s \approx 0.975$ and $r \approx 0.05$ marks the boundary between the two regimes, although it becomes much less sharp as $n_s$ increases. It is important to note that the density of points on these planes is controlled by the assumed priors on the distributions of potential parameters when drawing samples -- what matters here is the presence of even a single non-fine-tuned point with small $r$.
Thus the conclusion that large $r$ is correlated with $n_s$ for untuned models reached by applying BST is driven by its implicit assumption that the {\sl endpoint of inflation} is fixed by the same parameters that determine the shape of the potential at $\phi_{60}$. 

This is more clearly seen in Figure~\ref{boyle01graph} which selects out only the points which are not fine-tuned by the BST definition, i.e. the region for which $Z_\eta \leq 1$. The axes of this figure include the  $3 \sigma$ range of $n_s$ allowed by the most recent cosmological data, and values of $r$ which encompass both those accessible to next generation experiments ($r \sim 0.01$) as well as those which are effectively undetectable ($r \sim 10^{-3} - 10^{-5}$). 
The left panel of this figure, which reproduces the BST results for the full observationally-allowed range, shows that there exist natural models with $r<10^{-2}$ at the redder end of this range.  In BST's analysis, this region extends  to $n_s \sim 0.98$, while on our graph the region extends to $n_s \sim 0.99$, due to our use of $\phi$ derivatives to compute $Z_\eta$, rather than $N$ derivatives. In absolute terms, the difference in $n_s$ is small, but the relative breaking of scale invariance differs by a factor of 2, which is a non-trivial distinction and is driven by a small and not unreasonable change in the specification of the fine-tuning criterion.  

We also investigated the consequences of imposing the full set of BST selection conditions but adding higher order terms to the potential. With a $\phi^6$ term we reproduce the results proposed by BST on qualitative grounds. The only significant new region with $Z_\eta<2$ was within the formerly blank zone in the middle of the ``arms'' seen in the plots, and did not contain points with $r<10^{-2}$. 
Note that these ``forbidden zones'' seen in the plots are not due to under-sampling, but arise due to our chosen restrictions on the functional form and shape of the potential. If $n_s$ and $r$ are observed within these zones, inflation cannot be described by a purely quartic potential, as discussed in Appendix \ref{forbidden}.

\subsection{Initial Conditions Fine-tuning Criterion}\label{sec:Initialresults}

\begin{figure*}[!htp]
\includegraphics[scale=0.175]{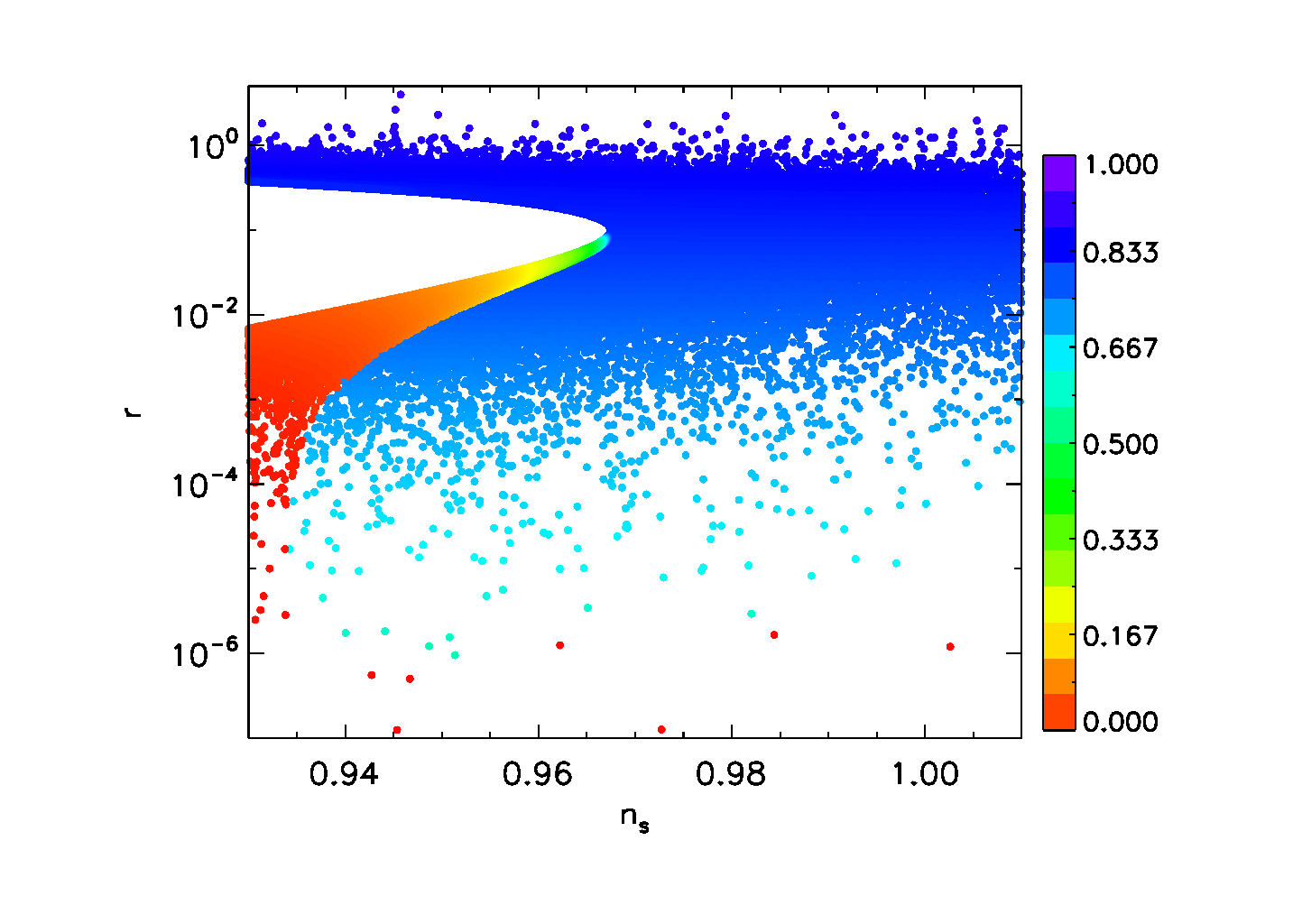} \hfill
\includegraphics[scale=0.175]{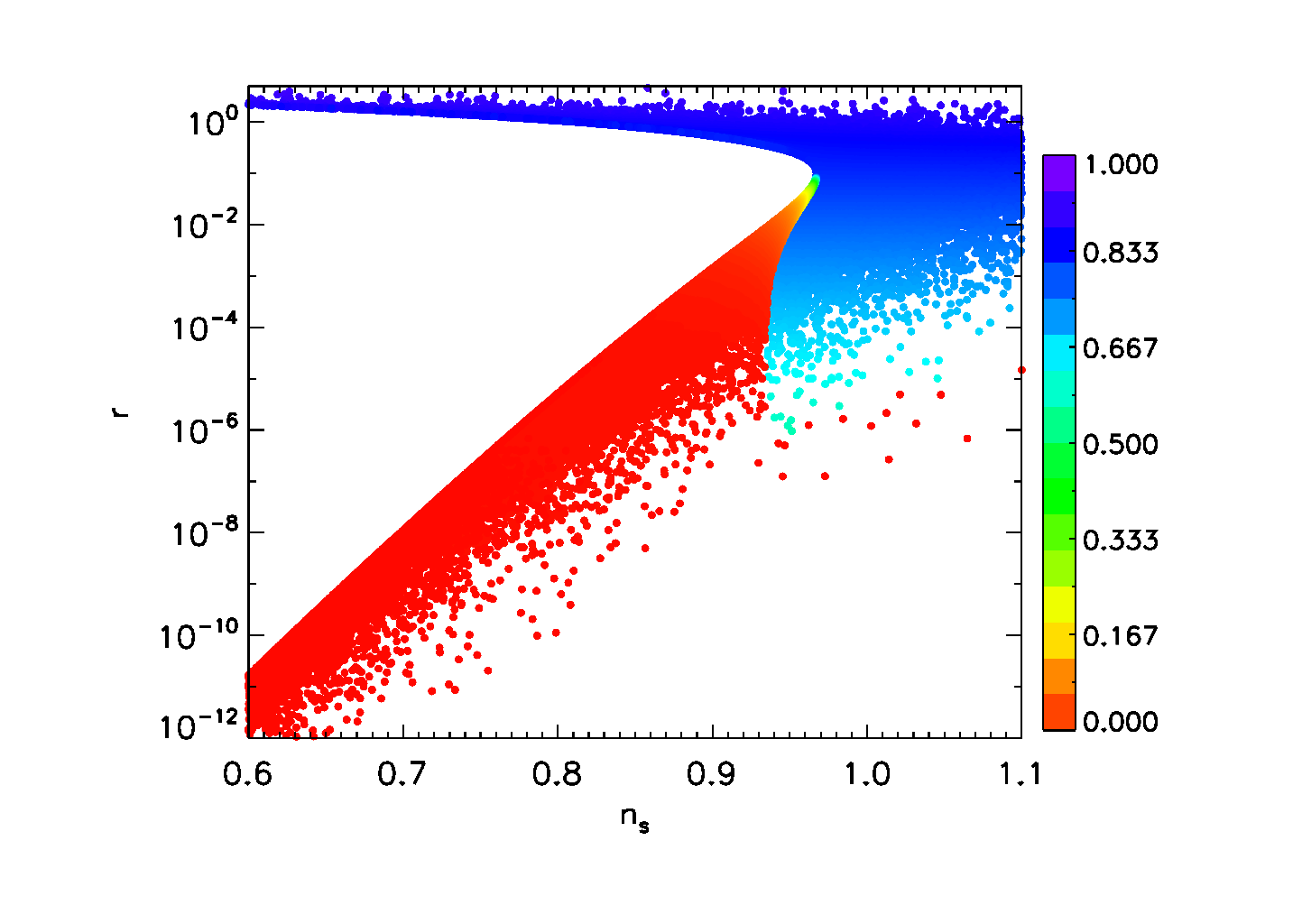}
\caption{The initial conditions fine-tuning criterion for the same potentials as shown in Figure~\ref{boylegraph}. The color scale shows the value of $R_i$ in the $(n_s,r)$ plane. Points with a lower $R_i$ are more fine-tuned. (Left) selected region showing observationally relevant values of $n_s$. (Right) an extended range.  See Appendix \ref{forbidden} for an explanation of the ``forbidden zone'' in the middle.} \label{rigraph}
\end{figure*}

Figure~\ref{rigraph} shows the equivalent plots to Figure~\ref{boylegraph}, for the initial conditions fine-tuning criterion. The potentials used to generate the plots are the same as Figure~\ref{boylegraph}. A model with smaller $R_i$ is more fine-tuned. 

We see two distinct regions; a ``head'' where $n_s\geq 0.96$, which has relatively high $R_i$, and a ``tail'' with relatively low $R_i$. In the head region, failed trajectories are dominated by overshoots, while in the tail the potential contains an extremum in the region of interest. Here, failed trajectories are dominated by those which originate on the far side of the maximum and have insufficient kinetic energy to pass over it. These potentials can appear natural in the BST criterion, because the feature occurs more than $60$ $e$-folds from the end of inflation and does not contribute to $Z_\eta$. Figure~\ref{rigraph} shows that the whole extended tail which the BST criterion considers natural has negligible $R_i$. This is clarified in Appendix \ref{forbidden}; the only way to have both low $n_s$ and low $r$ is for $\phisxty$ to be near a maximum. 

The head contains models which resemble the monomial potentials of chaotic inflation. Such potentials were examined in, e.g., ref. \cite{Belinsky:1985}, and found to be largely insensitive to their initial conditions; our results support this conclusion. In these potentials, the failed trajectories are almost entirely overshoots. Note that $R_i$ decreases slightly with $r$. This correlation is not as strong as that between the head and the tail, but it is still significant. For a fixed perturbation amplitude, a larger $r$ implies a larger $V'$ at $\phi_{60}$, as the magnitude of $V$ is fixed by condition (\ref{cond:perts}). The system enters the slow-roll regime when $\Pi \sim V'$, so a larger $V'$ will mean an earlier entry into slow-roll, and a corresponding decrease in the amount of overshoot.

The head continues down to $r\approx 10^{-6}$, and potentials with smaller $r$ are very fine-tuned. $V'/V$ must be tiny to achieve such a small $r$, and there is thus a minimum or inflection point in the vicinity. These are fine-tuned, first because they are sensitive to overshoot, and second because the inflection point can be followed by a steep slope, in which $\epsilon > 1$, prematurely ending inflation. Note that ref. \cite{Underwood:2008dh} concludes that these potentials are insensitive to initial conditions in the context of brane inflation, due to the speed limit imposed by the non-canonical Dirac-Born-Infeld kinetic term.

\begin{table}[!htp]
\begin{center}
\begin{tabular}{|c|c|c|}
\hline 
Parameter & Case 1 & Case 2\\
\hline
$n_s$ & $0.952$ & $0.963$ \\
$r$ & $2.34 \times 10^{-2}$ & $3.65\times 10^{-3}$ \\
$Z_\eta$ & $1$ & $7$ \\
$R_i$ & $0.137$ & $0.781$ \\
$V_0$ & $8.19 \times 10^{-18}$ & $2.38 \times 10^{-17}$ \\
$A$ & $777.0$ & $148.8$ \\
$B$ & $55.6$ & $22.86$ \\
$\phi_{60}$ & $-11.4$ & $-7.959$ \\
$\phi_e$ & $-1.34$ & $-1.264$ \\
\hline
\end{tabular}
\end{center} 
\caption{Examples of potentials where the two fine-tuning criteria lead to different conclusions. Inflation ends at $\phi_e$, when $\epsilon=1$.}
\label{models}
\end{table}

\begin{figure*}[!htp]
\includegraphics[scale=0.175]{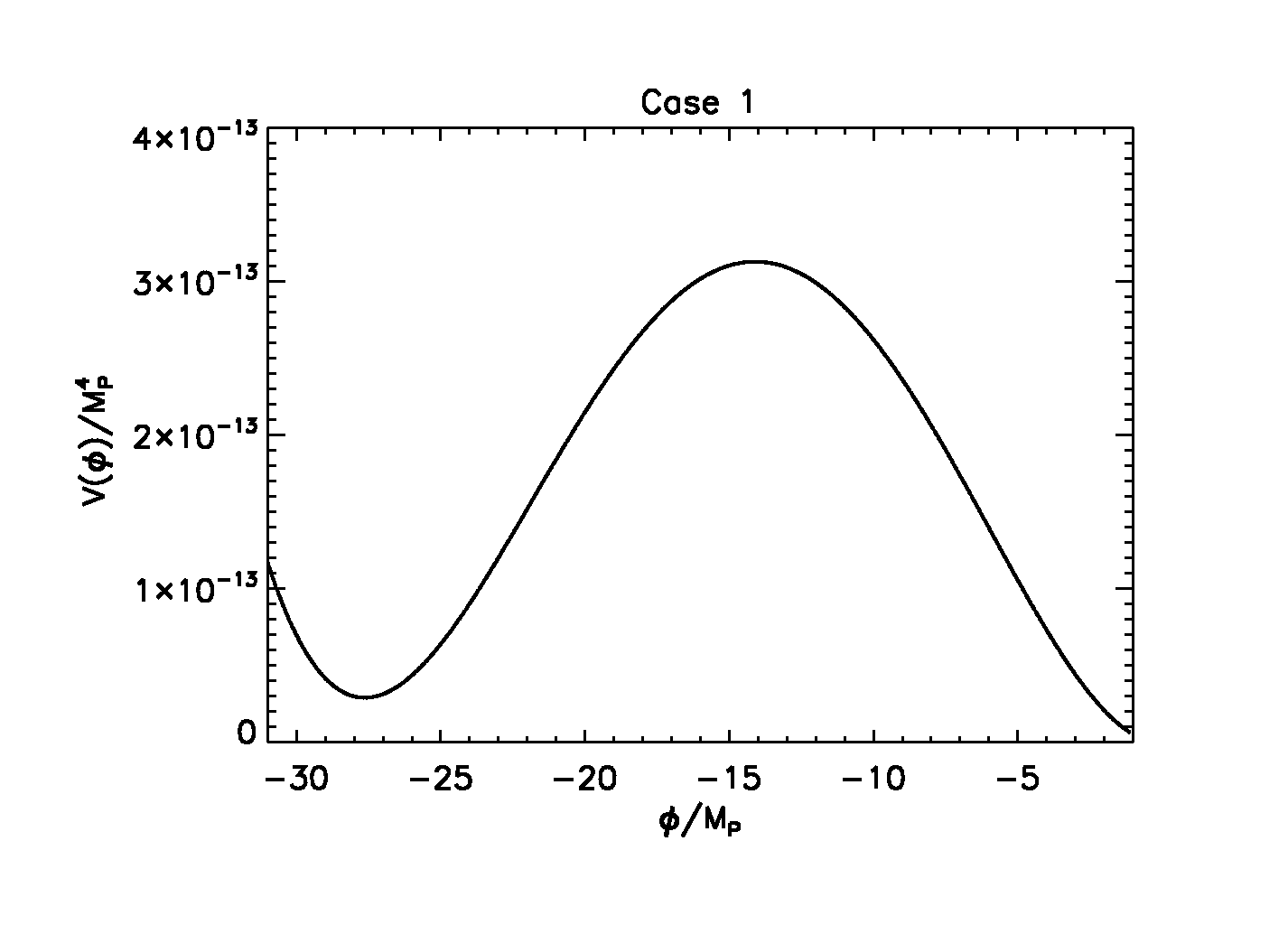} \hfill
\includegraphics[scale=0.175]{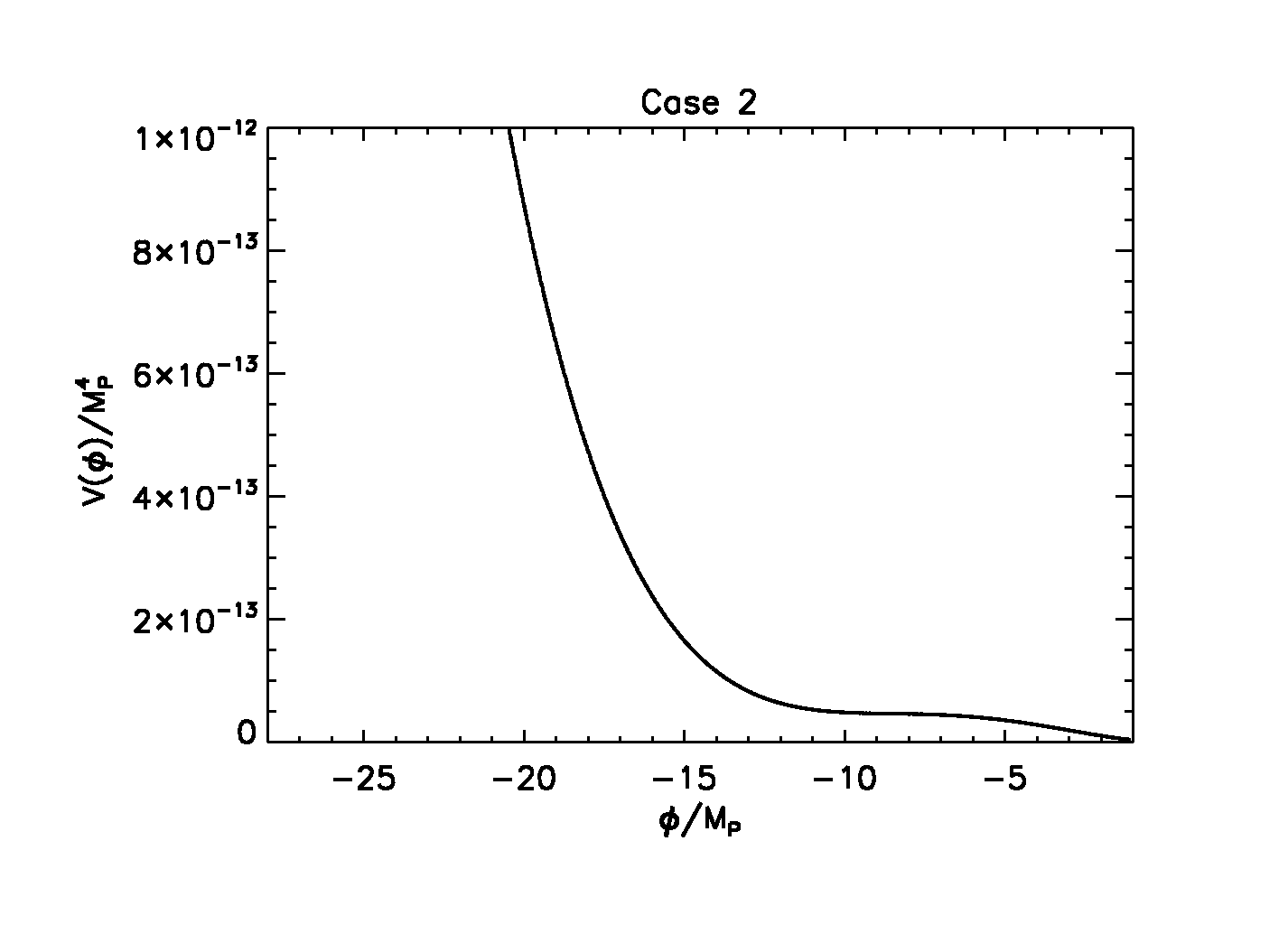} \\
\includegraphics[scale=0.175]{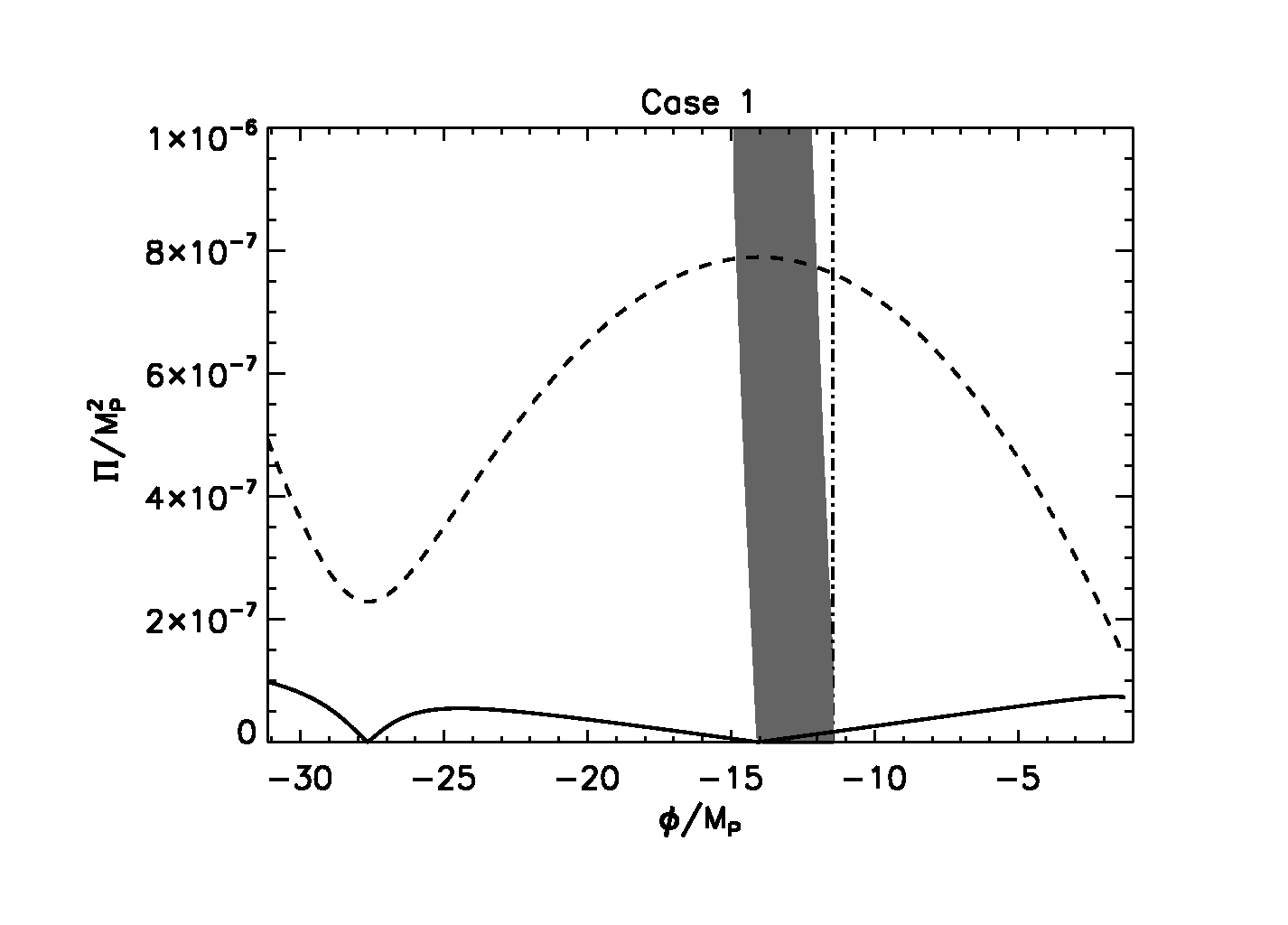}\hfill
\includegraphics[scale=0.175]{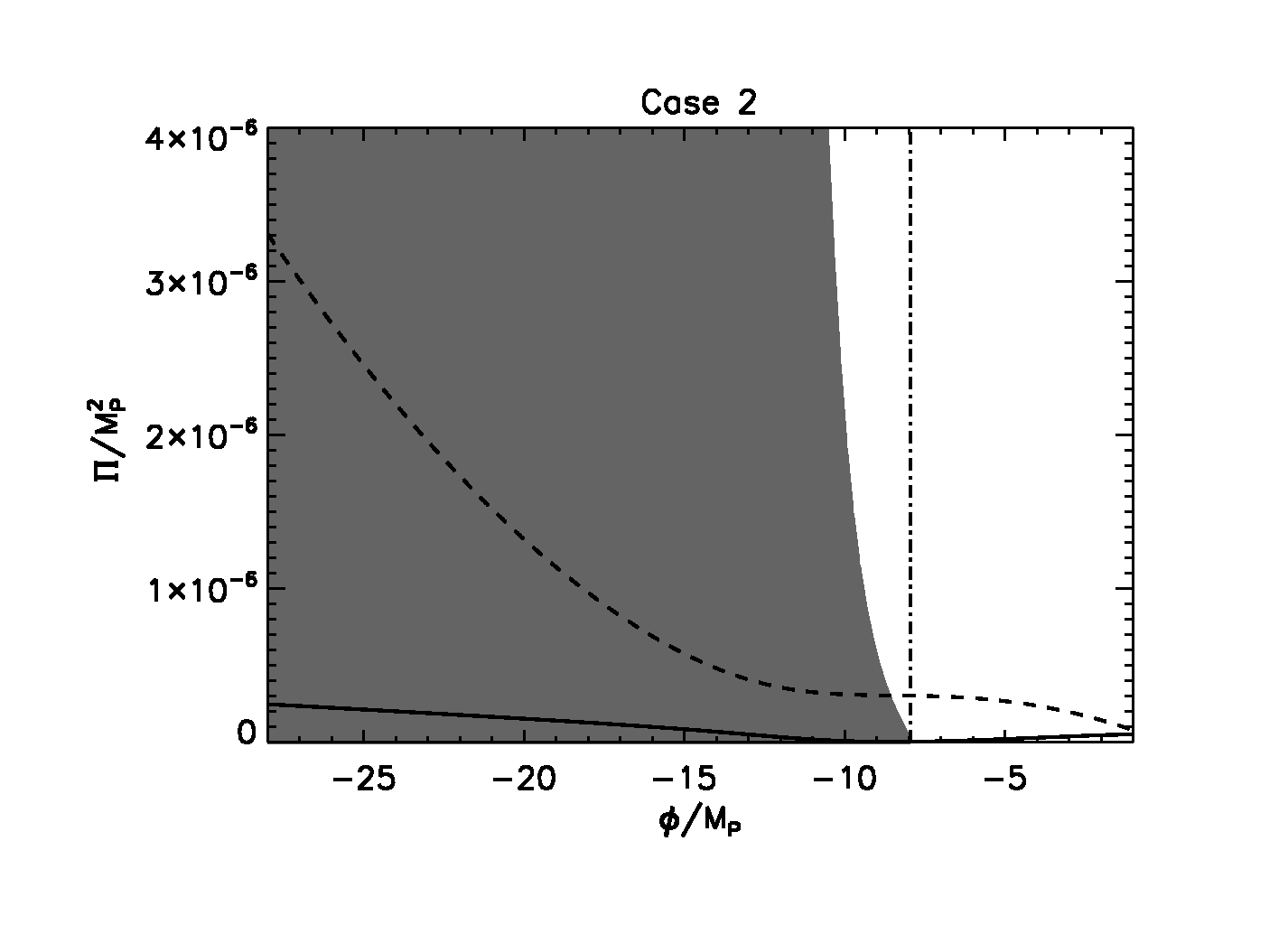}
\caption{(Top) The potentials for Case 1 (left) and Case 2 (right) -- see text and Table \ref{models}. The last $60$ $e$-folds of inflation take place between $-11.4 \lesssim \phi \lesssim -1.3$ for Case 1, and between $-7.96 \lesssim \phi \lesssim -1.26$ for Case 2. 
(Bottom) The gray region shows the portion of the initial conditions which lead to sufficient inflation in Case 1 (left) and Case 2 (right), while the white region shows a portion which does not. The dashed line shows the boundary between the slow-roll regime and the fast-roll regime, while the solid line shows the slow-roll attractor, and the dot-dashed line shows $\phi_{60}$. Only a subset of the $\Pi$ range tested is shown.\label{fig:cases}}
\end{figure*}

To highlight the complementary information provided by the two fine-tuning criteria, we present two examples in Table \ref{models} and Figure~\ref{fig:cases}. Case 1 has low $R_i$, but is not fine-tuned by BST's criterion. Case 2 is fine-tuned according to the BST criterion, but has high $R_i$. Case 2 has a flat plateau near $\phi_{60}$, which leads to the small value of $r$. Between the plateau and the end of inflation, there is a kink in the potential, during which $\eta$ changes sign, leading to the high BST fine-tuning.

Figure~\ref{fig:cases} shows that trajectories can travel through three major regimes \cite{Goldwirth:1992}: a ``fast-roll'' regime, where the kinetic energy dominates, a ``slow-roll'' regime, where inflation takes place, and an oscillatory regime. The oscillatory regime is entered as inflation ends, and does not concern us here\footnote{A WKB approximation shows that it begins when $9H^2 \approx 4V''$.}. Both fast and slow-roll regimes have $V' \ll 3H\Pi$. The condition for fast-roll is $\Pi^2 \gg V$, and for slow-roll, $\Pi^2 \ll V$. The equations of motion are, in the fast-roll regime:
\begin{equation}
\frac{d\Pi}{dN} \sim -3\Pi \qquad\frac{d\phi}{dN} \sim 1, \nonumber
\end{equation}
and in the slow-roll regime:
\begin{equation}
\frac{d\Pi}{dN} \sim -3\Pi \qquad \frac{d\phi}{dN} \sim \frac{\Pi}{\sqrt{V}}. \nonumber
\end{equation}

In both regimes, $\Pi$ is drawn strongly towards the slow-roll attractor. In slow-roll, $\phi$ rolls gently down the potential, while in fast-roll $\phi$ increases quickly, but little expansion occurs. If we return to our specific example potentials, trajectories with high initial kinetic energy decelerate rapidly towards the inflationary attractor. In Case 2, if the attractor is reached by a given trajectory, the inflaton will roll smoothly down the potential, yielding sufficient $e$-folds of inflation.  Overshoot trajectories appear when there is sufficient kinetic energy to prevent the system reaching the slow-roll phase with a negative enough $\phi$. 
In Case 1, the system must exit the fast-roll regime in a specific range of $\phi$ in order to achieve sufficient inflation. If slow-roll starts when $\phi < -15 M_P$, the inflaton will be on the wrong side of the maximum, unable to reach $\phi=0$. If it enters the slow-roll regime with $\phi \gg \phi_{60}$, it will overshoot. The initial $\Pi$ must therefore be such that the kinetic energy dissipates when $-11M_P < \phi < -15M_P$. This (relatively) narrow range causes the potential to appear fine-tuned.
 
\subsection{Common Inflationary Potentials}\label{sec:natresults}

In addition to the analysis of the quartic potential of equation (\ref{eq:eftpotential}), it is interesting to see what fine-tuning criteria say about standard potentials. We computed the values of $Z_\eta$ and $R_i$ for a set of these, choosing their free parameters to produce observables in agreement with the experimental bounds in Ref. \cite{Komatsu:2008hk}. 
The models we investigated are chaotic inflation \cite{Linde:1983}, new inflation \cite{Linde:1982, Guth:1982}, power-law inflation \cite{Lucchin:1985}, hybrid inflation \cite{Linde:1993cn}, natural inflation \cite{Freese:1990}, the Coleman-Weinberg potential \cite{Albrecht:1982,Linde:1983a} and two hilltop models \cite{Easther:2006qu,Efstathiou:2006}. Our results are shown in Table \ref{natural}. Our results agree with similar calculations performed for chaotic \cite{Belinsky:1985} new \cite{Goldwirth:1990} and natural \cite{Knox:1993} inflation. 

\begin{table}[!htp]
\begin{center}
\begin{tabular}{|c|c|c|c|c|c|c|}
\hline 
Model & Potential & $\mu$ & $n_s$ & $r$ & $ Z_\eta$ & $R_i$\\
\hline
\hline
Chaotic & $V_0\phi^\mu$ & $4$ & $0.949$ & $0.273$ & $0$ & $0.868$ \\
 & & $2$ & $0.966$ & $0.135$ & $0$ & $0.861$ \\
\hline
New & $V_0\left[(\phi/\mu)^2-1\right]^2$ & $15$ & $0.955$ & $0.0287$ & $1$ & $0.168$ \\
\hline
Power-law & $V_0\exp(-\phi\sqrt{2/\mu})$ & $120$ & $0.983$ & $0.133$ & $0$ & $1.0$ \\
\hline
Natural & $V_0\left[1+\cos\left(\phi/\mu\right)\right]$ & $10$\footnote{To compare our $\mu$ with the $f$ of ref. \cite{Freese:1990}, divide by $\sqrt{8\pi}$.} & $0.966$ & $0.0976$ & $0$ & $0.831$ \\
\hline
Hybrid\footnote{We are neglecting the second field here until $\phi=-\mu$, at which point we assume inflation ends abruptly. Thus, we are not applying BST selection conditions (\ref{cond:end})--(\ref{cond:min}) to this potential.}  & $V_0\left[1+\left(\phi/\mu\right)^2\right]$ & $1$ & $0.966$ & $0.137$ & $12$ & $0.864$ \\
& & $100$ &  $1.01$ & $9.0\times 10^{-4}$ & $0$ & $0.761$ \\
\hline
Hilltop\footnote{We are not applying BST selection condition (\ref{cond:min}) to this potential. Lower $r$ for this model comes at the cost of a $n_s$ that is too red compared to observations.}  & $V_0\left[1-\left(\phi/\mu\right)^2\right]$ & $10$ & $0.954$ & $0.011$ & $0$ & $0.010$ \\
& & $8.8$ & $0.945$ & $0.007$ & $0$ & $0.06$ \\
\hline
Hilltop\footnote{We are not applying BST selection condition (\ref{cond:min}) to this potential.}  
 & $V_0\left[1-\left(\phi/\mu\right)^4\right]$ & $4$ & $0.954$ & $2.2\times 10^{-4}$ & $1$ & $0.001$ \\
 & &  $20$ & $0.970$ & $0.018$ & $0$ & $0.60$ \\
\hline
Stabilized& $V_0[C-(\phi/\mu)^4$& $0.5$ & $0.955$ & $1.3\times 10^{-6}$ & $1$ & $0.00$ \\
Hilltop &   $\quad  + 0.01(\phi/\mu)^8]$ & $6$ & $0.963$ & $0.012$ & $3$ & $0.344$ \\
\hline
Coleman- & $V_0[1/4+\left(\phi/\mu\right)^4\times$ & $8$ & $0.947$ & $0.0013$ & $10$ & $0.06$ \\
Weinberg & $\left(\ln\left(\phi/\mu\right)-1/4\right)]$ &  $30$ & $0.967$ & $0.057$ & $1$ & $0.83$ \\ 
\hline
\end{tabular}
\caption{Fine-tuning criteria for specific potentials.}
\label{natural}
\end{center}
\end{table}

  For chaotic inflation (and power-law / natural inflation, which have  similar dynamical properties), all the fine-tuning criteria agree: they are ``natural'' in terms of symmetry considerations, turning points, and insensitivity to initial conditions. New inflation and hilltop type models are sensitive to their initial conditions in a way that is not accounted for by the BST fine-tuning criterion.
For parameter choices where the log term is significant, Coleman-Weinberg potentials appear fine-tuned in both criteria as the $\log \phi$ term generates a high $Z_\eta$, while their sensitivity to initial conditions is well-known \cite{Goldwirth:1992}. For a sufficiently large value of $\mu$, the log term becomes unimportant and the potential resembles that of chaotic inflation. However, these potentials may be technically natural, in the sense that they are stable against (further) loop corrections.   The Coleman-Weinberg potential thus has one parameter, and in the absence of a microscopic theory, no value of $\mu$ is favored over any other. However, both $Z_\eta$ and $R_i$ change significantly  between  $\mu=8$ and $\mu = 30$. Thus the tuning criteria will penalize the potential for not being in the ``chaotic'' regime, even though the underlying physics has not changed.

Hybrid inflation, which can be highly fine-tuned by the BST criterion, in addition to lacking a deterministic end to inflation (since the endpoint of inflation is fixed by the coupling between $\phi$ and a second field, which is not included in this potential), is very insensitive to initial conditions.   In these cases, the BST trend is reversed; models with low $r$ are less fine-tuned. A hybrid model with $\mu \gg 1$ has $V \sim V_0$, $\eta \propto 1/\mu^2$, $Z_\eta=0$, $n_s \sim 1$, and unobservable $r$. Conversely, the hybrid case with $\mu \ll 1$ resembles chaotic $\phi^2$ inflation, and hence has $Z_\eta=0$. The intermediate regime $\mu \sim 1$ balances the two asymptotic forms of $V$ and hence has high $Z_\eta$, even though $r$ is large and $n_s \sim 0.96$.

Finally, we consider a potential which is stabilized by a higher order term, but is constructed to meet all five of the BST conditions,
\begin{equation}
V=V_0\left[C-\left(\frac{\phi}{\mu}\right)^4+\delta\left(\frac{\phi}{\mu}\right)^8\right] \, .
\end{equation}
The value of $C$ is fixed by condition (\ref{cond:min}), with $V=0$ at the minimum, and $\delta$ is assumed to be small ($10^{-2}$).\footnote{In the original formulation of the BST criterion, polynomial potentials of degree $n>4$ have $Z_{\rm total} = Z_\eta + n-4$,  but we will work  with a $Z_\eta$ derived solely from the shape of the potential.}
For $\mu<0.8$ this potential has $Z_\eta=1$, together with a very low value of $r$. It provides a counter-example to BST's claim that natural potentials require $r>10^{-2}$. It is, however, very fine-tuned with respect to initial conditions. Higher values of $\mu$ do not involve fine-tuned initial conditions but have a larger $r$, as well as a large $Z_\eta$.

\section{Conclusions}\label{sec:conclusions}

We have investigated whether, given the observed limits on the scalar tilt $n_s$, it is possible to claim that certain ranges of values for the tensor-scalar ratio $r$ are ``fine-tuned'' in a way that is independent of the definition of fine-tuning. We find that, on the contrary, different criteria are sensitive to different types of fine-tuning, and hence can give differing answers about the naturalness of a given potential. The BST criterion, based on the number of unnecessary ``features" in the potential, tells us that models with $r<10^{-2}$ are fine-tuned for $n_s > 0.95$, while an alternative criterion based on sensitivity to initial conditions shows that models with $n_s<0.94$ and $r<10^{-3}$ are fine-tuned, but with $n_s>0.94$ there are ``natural'' models with $r \sim 10^{-6}$. 

Moreover, BST's conclusions are not robust against relatively minor modifications to their criteria.  In particular, the claim that models with $n_s>0.98$ are fine-tuned is softened by a small technical change in the fine-tuning criterion $Z_\eta$: our upper limit on natural values of the spectral index is $n_s <0.99$.  This may seem like a fine distinction, but recall that the breaking of scale invariance is measured by $n_s-1$, and thus differs by a factor of 2.  Moreover, Planck is expected to measure  $n_s$ with a 1-$\sigma$ error of $0.0045$ \cite{Planck} and  $n_s = 0.99$ cannot be easily distinguished from $n_s = 1$, but $n_s=0.98$ could be detected with considerable confidence.  Likewise,  BST's  conclusion that $n_s>0.95$ implies $r>10^{-2}$ in the absence of fine-tuning only holds if the potential evolves smoothly to an analytic minimum with $V(0) \approx 0$. This is equivalent to requiring that the parameters fixing the shape of the potential as cosmological perturbations are laid down also govern the shape of the potential as inflation ends.   While it is not inherently unreasonable, this restriction immediately excludes both hybrid inflation and many stringy constructions, even though their underlying potentials may be technically natural from a particle physics perspective. Pending a better understanding of physics at the energy scale of inflation, this condition therefore seems overly restrictive.
 
Our alternative criterion is based on the sensitivity of a particular potential to the inflaton's initial conditions. It shows that potentials with $n_s<0.94$ and $r<10^{-3}$ are very sensitive to their initial conditions, as are all models  with $r<10^{-6}$. Since inflation was originally designed to solve the initial condition problems of a hot big bang, one should certainly check that inflation itself does not require the universe to be in a special initial state.   For our ``scoring'' of the initial conditions problem, there is a small preference for models with higher $r$, but the {\em specific\/} results we find depend on the explicit construction of our fine-tuning criterion, so we do not see this as a robust conclusion. Rather, we emphasize that this criterion singles out a different set of potentials from the BST criterion, and neither set of potentials is a strict subset of the other.   The above conclusions apply to potentials that are well-described by a quartic potential. In Appendix \ref{potentialparams}, we explore the regions of parameter space  which yield untuned potentials. 
\begin{figure*}[!htp]
\includegraphics[scale=0.175]{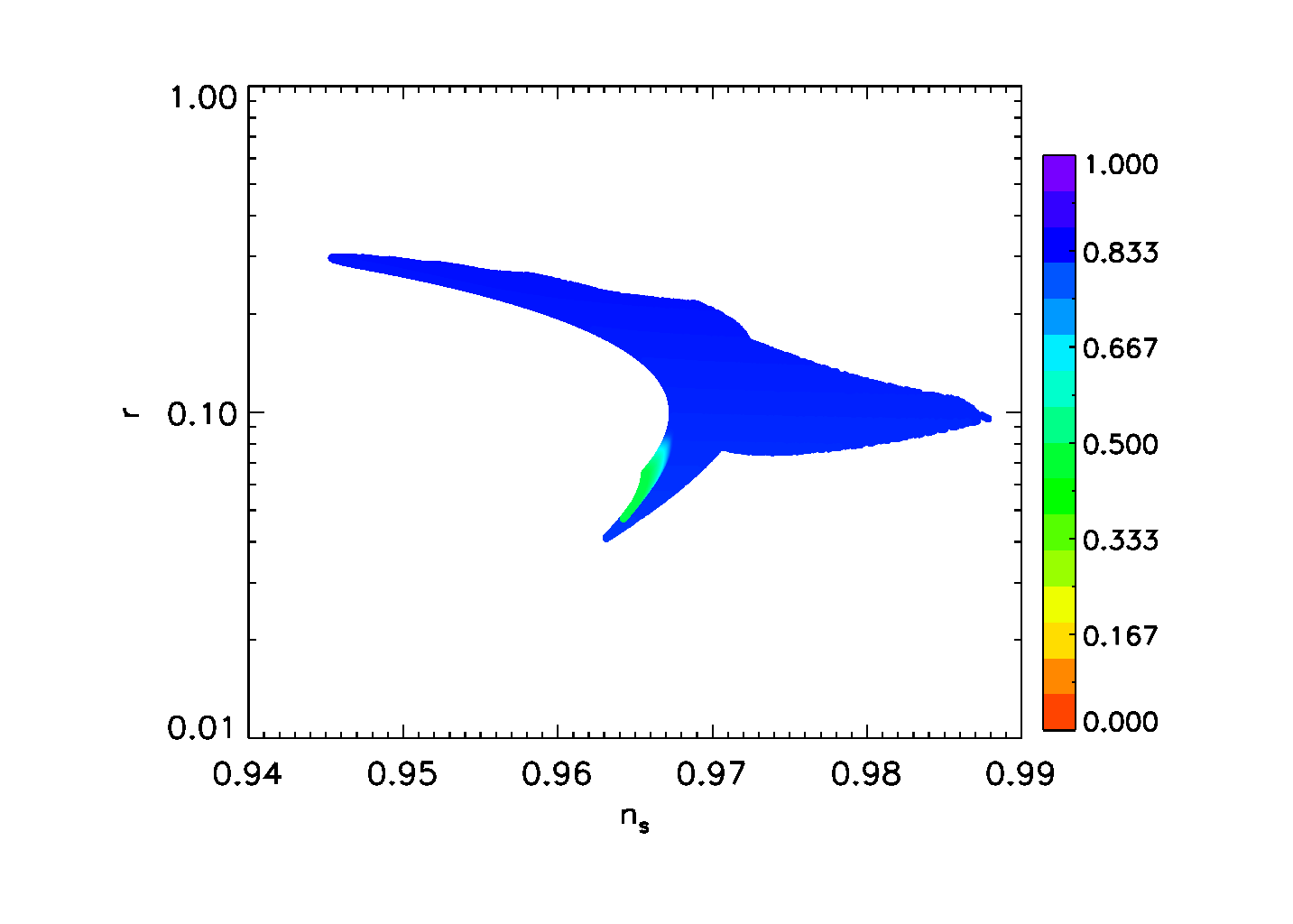}  \hfill
\includegraphics[scale=0.175]{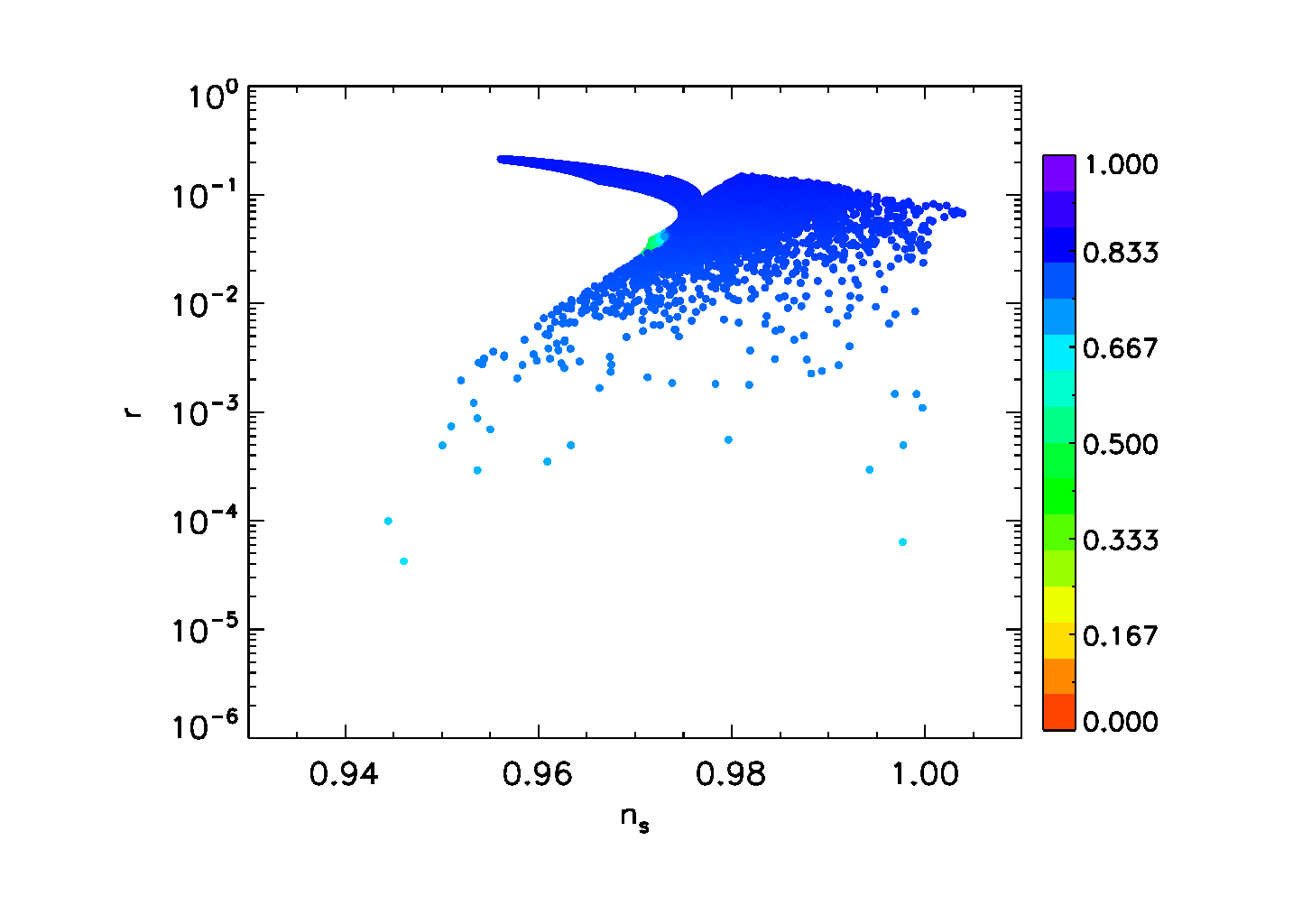} \\
\caption{(Top) The set of models which are deemed natural by both the BST and initial conditions criteria, i.e. $Z_\eta<2$ and $R_i>0.5$. The color bar shows $R_i$. 
(Left) Imposing all five BST conditions. (Right) Neglecting selection conditions (\ref{cond:min}) and (\ref{cond:meta}). 
Potentials which have low $r \sim 10^{-4}$ and $n_s \sim 0.95$ generically contain a flat plateau terminating in a drop, similar to Case 2 in Figure~\ref{fig:cases}.} \label{combinedft}
\end{figure*}

We have shown that conclusions drawn from   BST-like criteria depend on the explicit assumptions they encode.   We also discussed a second type of fine-tuning -- sensitivity to the initial field configuration -- which is distinct from that described of BST.   One may restrict attention to models deemed acceptable  by  {\em both\/} criteria. The left panel of Figure~\ref{combinedft} indicates that, at face value, this would lead one to conclude that the ``natural'' lower limit on the tensor amplitude is   $r \gtrsim 0.04$, and thus well within the range that is experimentally accessible.   However, the right panel shows that this conclusion only holds if condition (\ref{cond:min}) is imposed.   Moreover, there are types of fine-tuning missed by our initial conditions criterion (e.g. initial inhomogeneity or spatial curvature); including these would further reduce the number of models for which inflation can be regarded as generic.

The BST criterion aims to characterize the fine-tuning of the {\it shape} of the potential.  The initial conditions criterion might, at face value, appear to measure only the sensitivity to an initially large
kinetic term. However, it is important to realize that it also encodes information about the functional form of the potential: for instance, a potential with a steep hill, a small plateau, and another steep downhill
stretch will have difficulty supporting inflation without carefully tuned initial conditions. Thus, $R_i$ penalizes the functional form of the potential, but in a different way to $Z_\eta$. Hence, Occam's razor cannot  characterize fine-tuning (even of a single property, such as the shape of the potential) in an unequivocal way. These ambiguities will remain after the arrival of data with sufficient precision to select a small region of the $(r, n_s)$ plane. If this region is disfavored according to a specific fine-tuning criterion, should we interpret this as evidence against inflation, or even (in a more limited sense) as evidence against a class of inflationary models? Absent a detailed understanding of the physics of inflation, the answer is unclear, since
we would not know whether the fine-tuning identified by that criterion is relevant to the underlying theory.

As is well known, simple single field models of inflation have a direct  correlation between the total excursion of the inflaton field  and the value of $r$.  If values of $r>10^{-3}$  are ruled out by experiment, this would also eliminate all  single field models with $\Delta \phi>M_P$, with dramatic consequences for inflationary model building,  as well as for alternative theories of the early universe \cite{Khoury:2001bz,Boyle:2003km}. Many proposed inflationary models predict signals in this range but only a handful of models live in the lower part, the so-called ``tensor desert'' at  $10^{-3}<r<10^{-2}$ \cite{Smith:2008pf,Alabidi:2006fu, Efstathiou:2006}. Conversely, many other models have $r \ll 10^{-3}$.   Consequently, it is clear that the inflationary paradigm would not be ruled out by a failure to detect primordial tensors, although  a large and important class of models would be conclusively eliminated.  

Our experience with different fine-tuning criteria suggests that it is likely to be impossible to construct a set of criteria which defines a ``generic inflationary model'' that is simultaneously (a) robust against small changes, (b) does not eliminate apparently reasonable models from consideration, and (c)  leads to a definitive conclusion for the likely value of the tensor-scalar ratio $r$.
Rather, the value of analyses like that of BST lies in the extent to which they identify a category of inflationary models which would be falsified if future experiments fail to detect tensors. However, determining that $r\lesssim 0.001$ -- or any other value likely to be experimentally accessible -- cannot be construed as a definitive test of inflation.

\appendix

\section{The Forbidden Zone} \label{forbidden}

The blank ``forbidden zone'' in Figures~\ref{boylegraph}-\ref{rigraph} is not an artifact of the selection priors; it is inaccessible for quartic potentials. Higher order potentials have a smaller forbidden zone, and cubic potentials a larger one (see Figure~\ref{cubicgraph}). In order to acquire some intuition about why this is the case, we shall now attempt to minimize $n_s-1$ for fixed $r$. In what follows, $M_P=1$. Recall that:\label{forbid}
\begin{eqnarray}
n_s &=& 1+2\eta-6\epsilon \\ 
&\approx& 1+2\frac{V''}{V} \qquad\mathrm{(for\ small}\ r\mathrm{).}
\end{eqnarray} 
Thus, $r < 10^{-2}$ implies $\epsilon \lesssim 10^{-3}$, so if in addition $n_s \sim 0.9$, $V''$ is negative and reasonably large. Now:
\begin{equation}
V''(\phi_{60}) < 0 \implies 12\phi_{60}^2+6B\phi_{60}+2A <0\,.
\end{equation}
Since $\phi < 0 < A$, this implies $B>0$. If $V'$ is small and $V''$ is large, a zero of $V'$ must be nearby. So, for large $n_s-1$ and small $r$, $\phisxty$ will be fairly near a maximum of the potential ($\phim$)\footnote{If we desired both $\eta$ and $\epsilon$ to be small, we could achieve this with a flat potential and no necessity for a maximum. This conclusion is borne out by Figure~\ref{rigraph}, where the high $R_i$ for points with $n_s \sim 1$ and small $r$ show that no maxima are nearby.}. Furthermore, $\phisxty > \phim$, because $N\to \infty$ as $\phisxty \to \phim$. Therefore, we shall consider all relevant quantities evaluated at $\phim$, Taylor-expanding around this point when the description is inadequate.

If a maximum exists, it occurs at:
\begin{eqnarray}
\phim = \frac{-3B + \sqrt{9B^2 - 32A}}{8}\,.
\end{eqnarray}
For $A>0$, this is a strictly decreasing function of $B$. Recall that $B^2 < 4A$, due to condition (\ref{cond:meta}), so
\begin{equation}
\phim > -\frac{\sqrt{A}}{2}.
\end{equation}
A lower bound on $V''$, therefore, is:
\begin{eqnarray}
V'' &>& V_0(12\phim^2+6B\phim+2A) \nonumber\\
 	&>& V_0(12\phim^2+12\sqrt{A}\phim+2A) \nonumber\\
 	&>& V_0(3A -6A+2A)  \nonumber\\
 	&>& -V_0A\,.
\end{eqnarray}
Since this minimal $V''$ is at a maximum of $V$, it also locally minimizes $\eta$, so:
\begin{eqnarray}
V &=& \frac{V_0A^2}{16} \nonumber\\
\eta &>& -\frac{16}{A}\,.
\end{eqnarray}
We cannot simply evaluate $V'$ at $\phim$, because this would tell us that $r=0$. The leading order of the Taylor expansion is significant here. We compute it via
\begin{eqnarray}
V' &\approx& V'(\phim) + V''(\phim)\frac{(\phisxty-\phim)}{2} \nonumber\\
	&\approx& (\phisxty-\phim)\frac{V_0A}{2} \\
\implies \epsilon &\approx& \frac{(\phisxty-\phim)^2}{2}\left(\frac{16}{A}\right)^2 \\
\implies r &\approx& \frac{2048}{A^2}\left(\phisxty+\frac{\sqrt{A}}{2}\right)^2\,.
\end{eqnarray}
Here our simple analytic approximations can go no further, because $\phisxty$ is a non-linear function of both $A$ and $B$. However, when $r \sim 10^{-2}$, numerical calculations show that $\phisxty-\phim \sim 1$. Case 1 (see Figure~\ref{fig:cases} and Table \ref{models}) has $\phisxty-\phim \approx 2$. With this final approximation, we obtain:
\begin{equation}
n_s \gtrsim 1-\sqrt{\frac{r}{2}}\,.
\end{equation}
Even when the final rough fit is no longer valid, it can be seen that on the lower arm of Figure~\ref{boylegraph}, reaching low $n_s$ requires a very low $r$, and  the potential which minimizes $n_s$ for given $r$ will contain a maximum.

For large $r$, this lower bound on $n_s$ ceases to hold, both because $\epsilon$ begins to become significant in the expression for $n_s$ and because $\phisxty$ need no longer be in proximity to a maximum. These models tend to involve a large variation in $\phi$ during inflation, $\eta \ll \epsilon$, and positive $V''$. Hence a better approximation for the minimum $n_s$ is: 
\begin{equation}
n_s \gtrsim 1 - \frac{3r}{8}\,.
\end{equation}
Matching these two limits together, we obtain a rough description of the forbidden zone.

\section{Potential Parameters}\label{potentialparams}

\begin{figure*}
\includegraphics[scale=0.175]{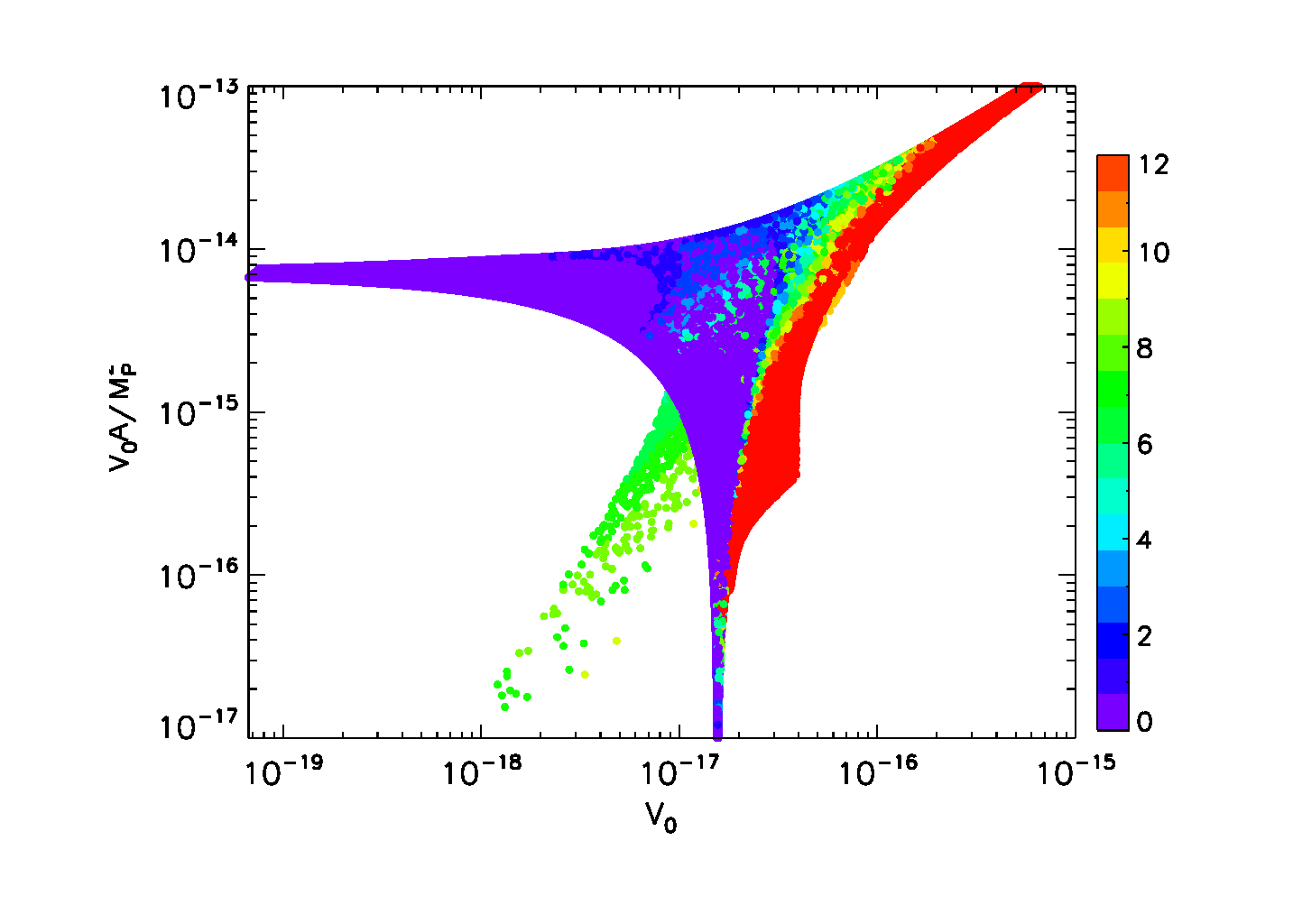} \hfill
\includegraphics[scale=0.175]{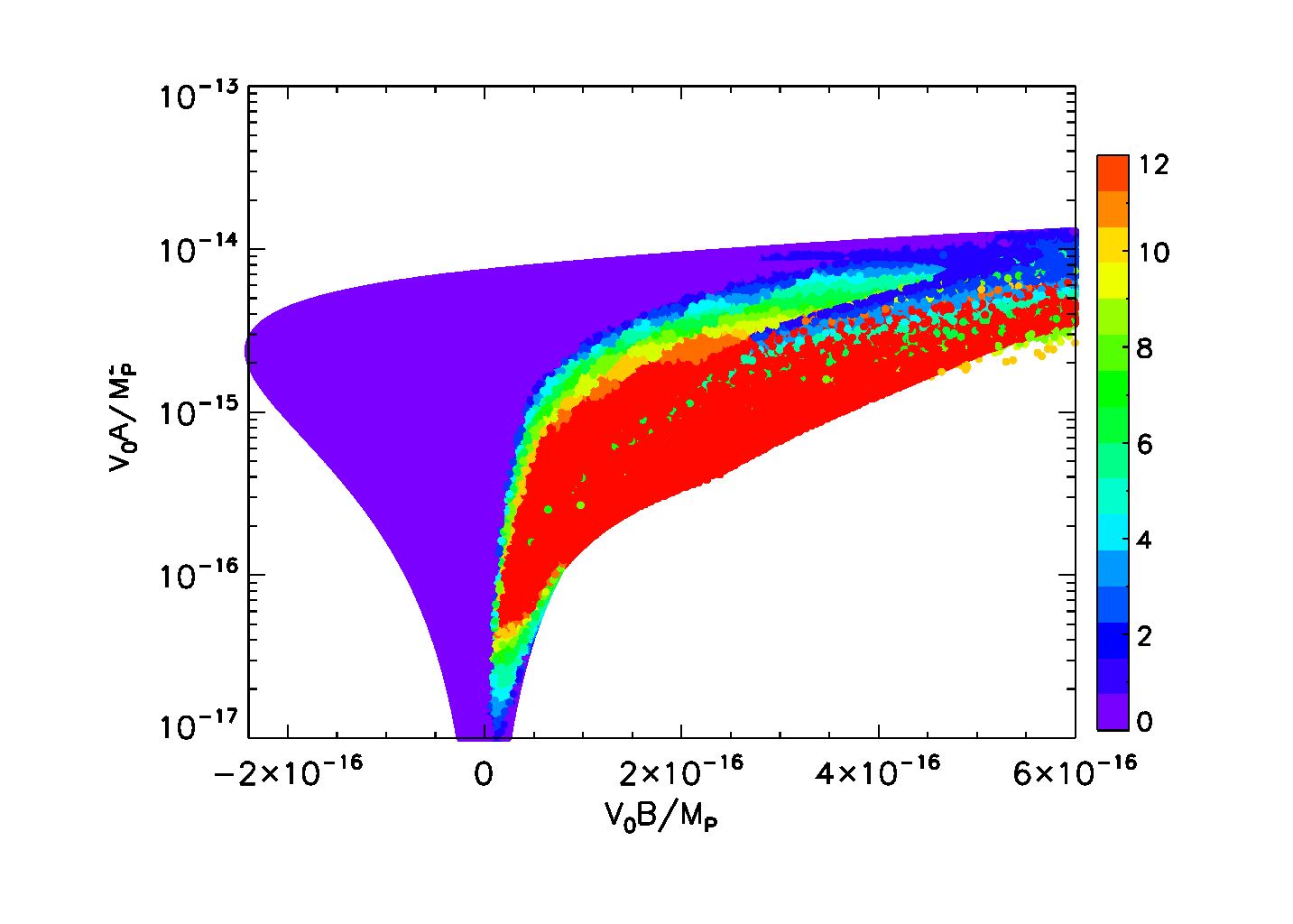}\\
\includegraphics[scale=0.175]{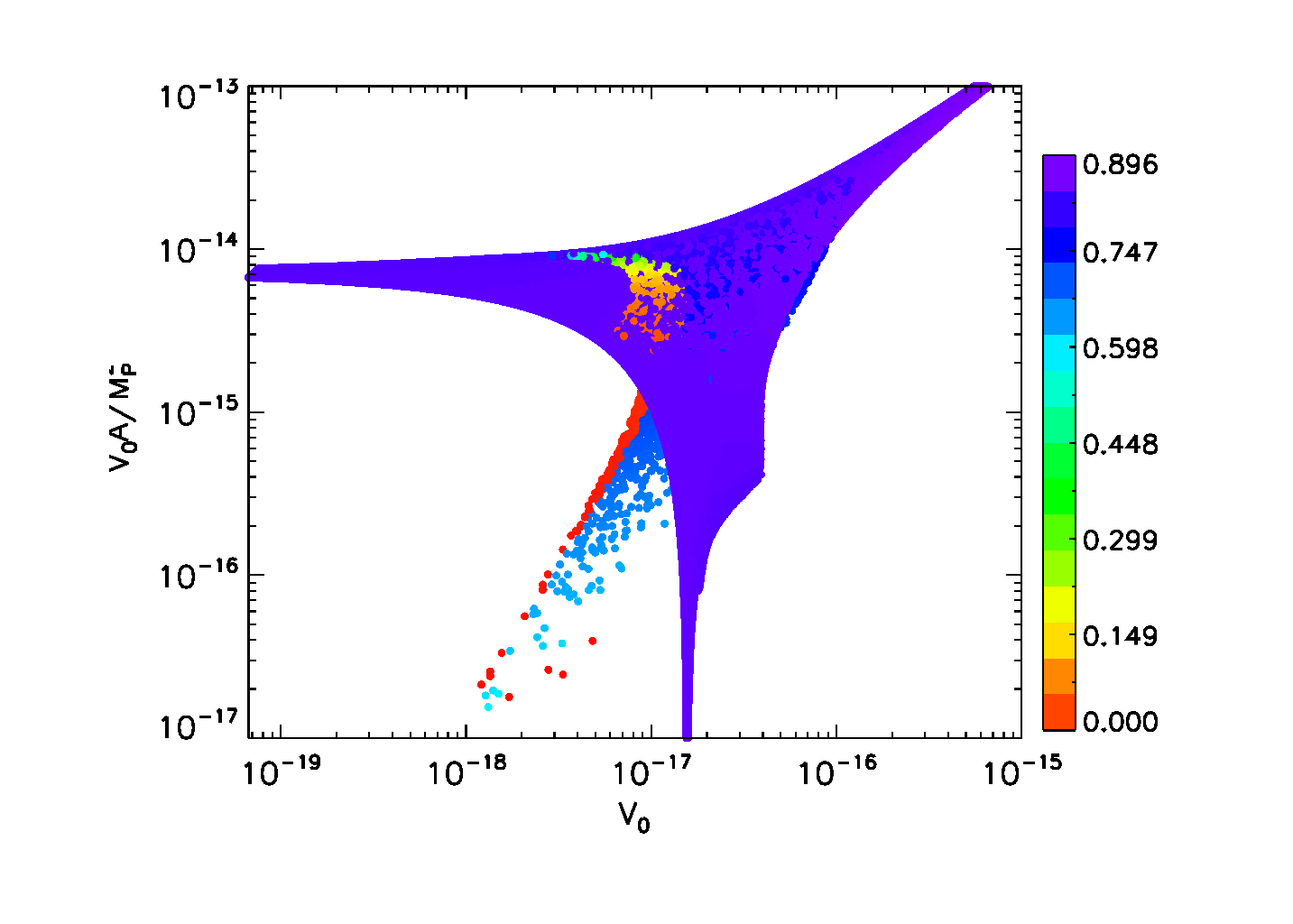} \hfill
\includegraphics[scale=0.175]{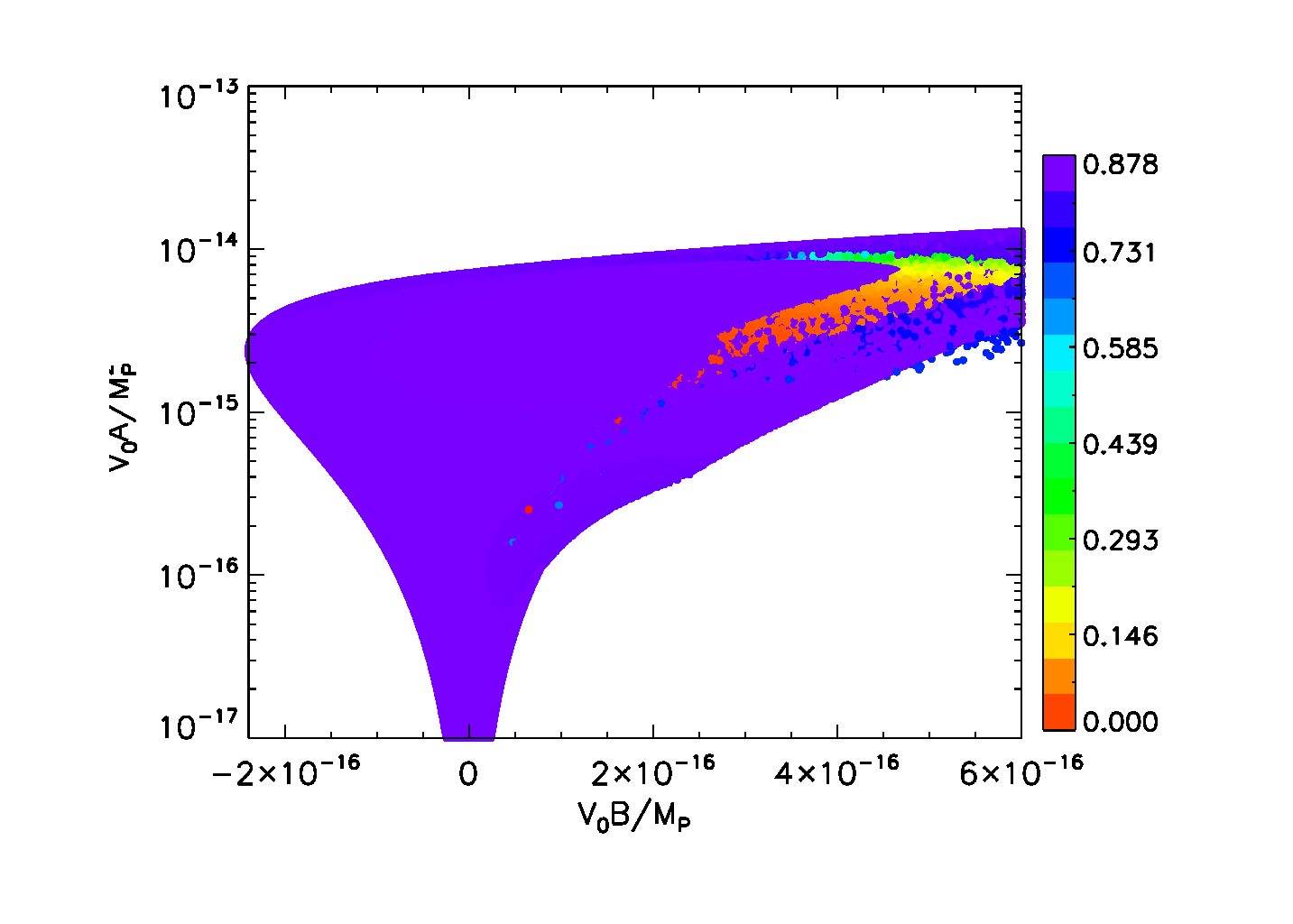}\\
\includegraphics[scale=0.175]{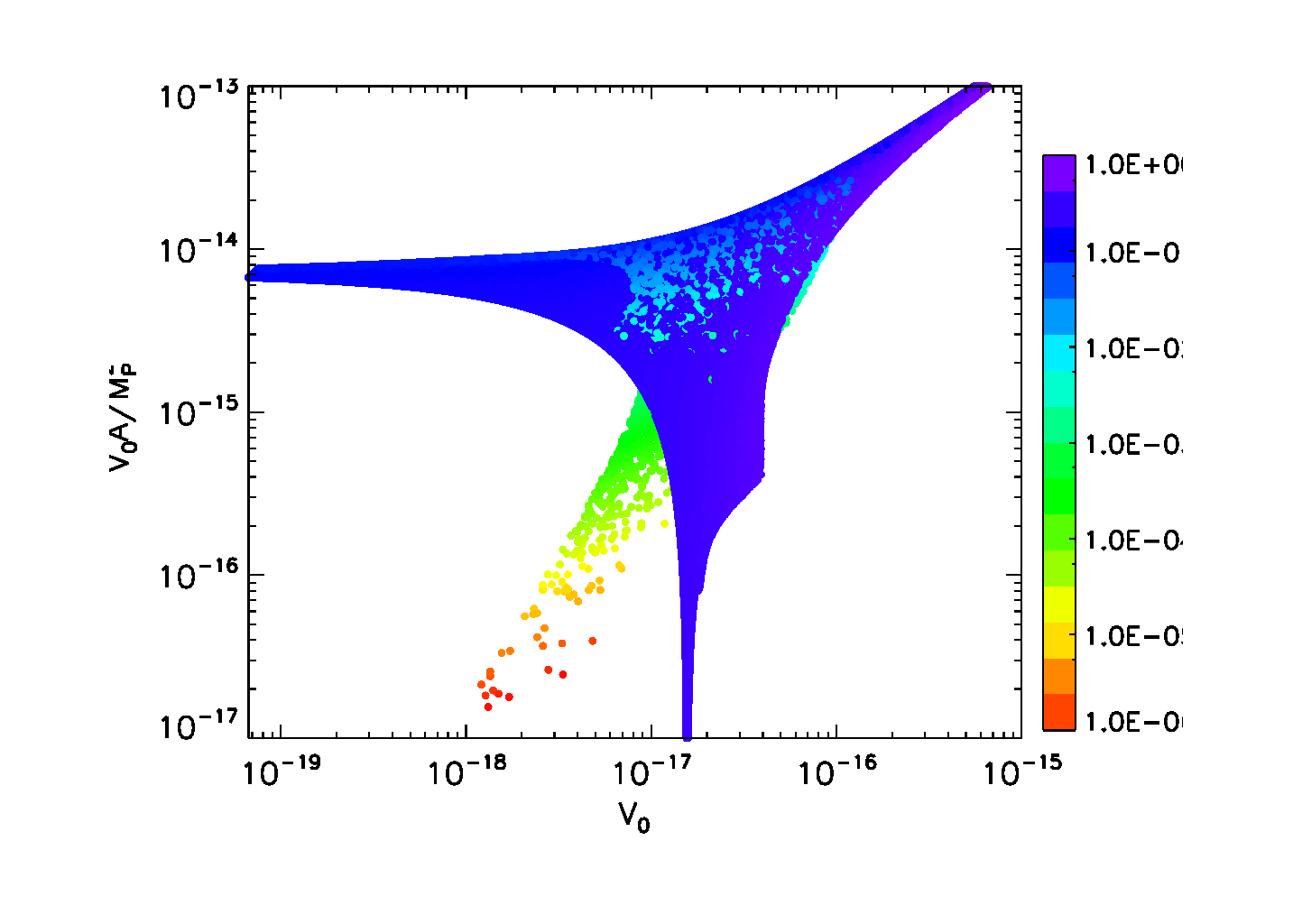}\hfill
\includegraphics[scale=0.175]{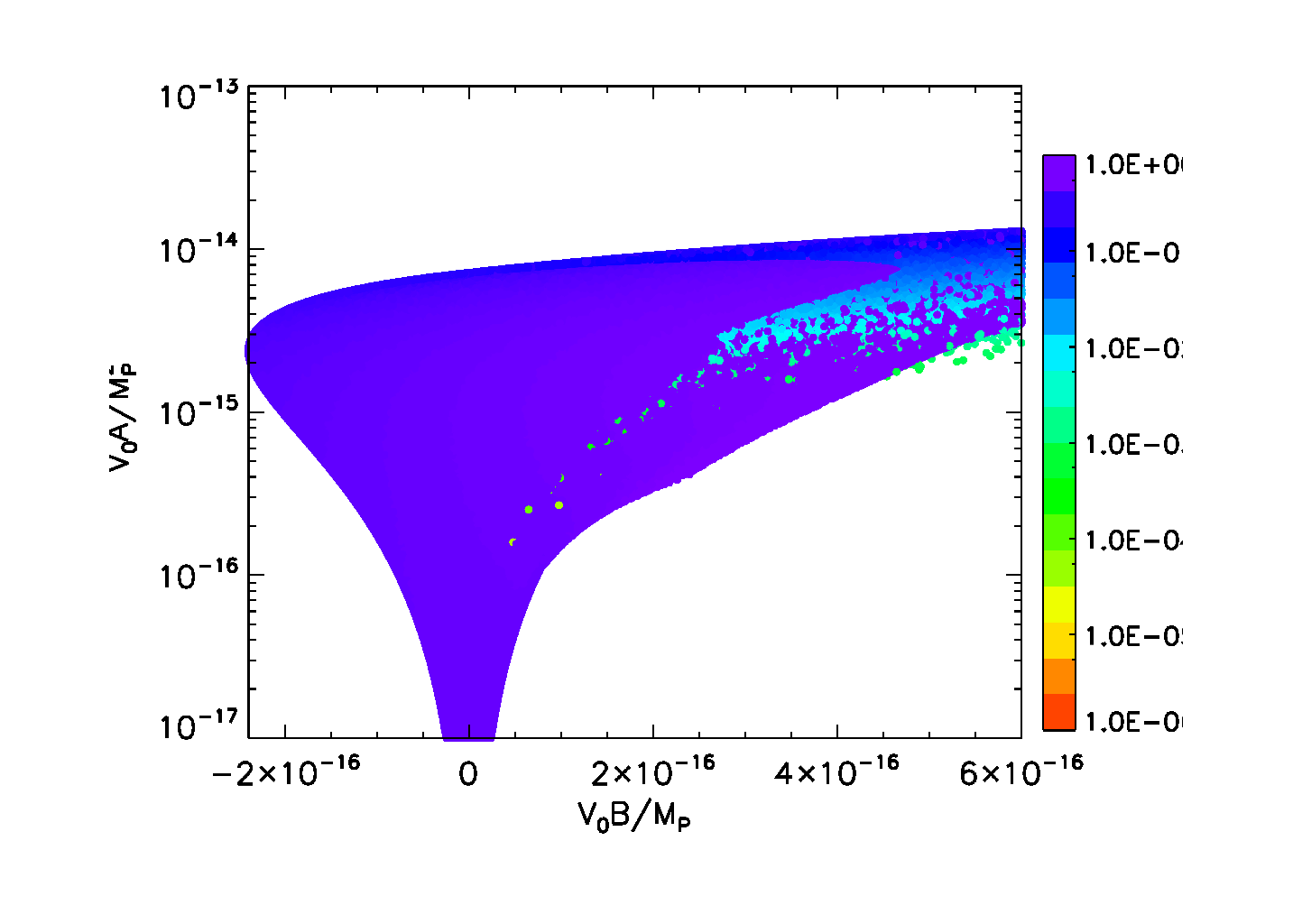}
\caption{(Left) The coefficient of $\phi^4$ is plotted against the coefficient of $\phi^2$. (Right) The coefficient of $\phi^3$ is plotted against the coefficient of $\phi^2$. Various quantities of interest are shown on a color scale: $Z_\eta$ (top), $R_i$ (middle) and $r$ (bottom). 
Only those points with $r>10^{-6}$ and $0.93 < n_s <1.02$ are shown. The potential is $V=V_0(\phi^4 + B\phi^3 +A\phi^2)$. In order to enhance clarity, these plots have been cropped to $10^{-17} < V_0A < 10^{-13}$. In addition, the right-hand plots are truncated at $V_0B = 8 \times 10^{-16}$; the points continue on to $V_0B \sim 10^{-11}$, becoming increasingly thinner, and with a gradually increasing $V_0A$, up to $10^{-11}$. The vertical line extending down to very small $V_0A$ on both plots is tending to $\lambda\phi^4$ inflation. 
The horizontal line at $V_0A \approx 10^{-14}$ corresponds to the $m^2 \phi^2$ inflation limit. The points in the truncated region are similar to those at the edges of each plot; those on the right have high $r$, $R_i$, and $Z_\eta$. .}\label{fig:potential}
\end{figure*}

Figure~\ref{fig:potential} shows how the space of potential parameters is weighted by the fine-tuning criteria and the resulting tensor-scalar ratio. When performing parameter estimation, we would normally place maximally uninformative priors upon these parameters. We plot $V_0B$ linearly, but $V_0$ and $V_0A$ are constrained to be positive, so we plot these logarithmically -- this mimics the flat prior that would be imposed on the former, and the logarithmic priors upon the latter. Because these graphs are projections of a volume, points with different values can overlap each other. $n_s$ is not shown, but is constrained to be $0.93 < n_s < 1.01$. 
The points on the left-hand plots with $10^{-14} < V_0A < 10^{-13}$ do not appear on the right-hand plots because they have cubic coefficients larger than those plotted. 
The quantity of empty space in these plots shows that the selection conditions have imposed a strong theoretical prior, but further work is needed to discover which of the conditions have had the strongest effect. BST deems that only the points which are color-coded in purple are not fine-tuned. We also see that a high $r$ seems a common occurrence (although by no means guaranteed).

\acknowledgments We thank Daniel Baumann, Asantha Cooray, Latham Boyle, Steven Gratton, William Kinney, David Seery, and Daniel Wesley for useful discussions. We also thank Daniel Baumann and Daniel Wesley for carefully proof-reading an earlier draft of this manuscript. SB is supported by STFC. HVP is supported in part by Marie Curie grant MIRG-CT-2007-203314 from the European Commission, and by a STFC Advanced Fellowship. RE is supported in part by the United States Department of Energy, grant DE-FG02-92ER-40704 and by an NSF Career Award PHY-0747868.

\bibliography{finetuning}

\end{document}